\documentclass[article]{JHEP3}

\usepackage{amsmath,epsfig}

\usepackage{amssymb,amsfonts}

\title{Type IIB Colliding Plane Waves}
\preprint{\hepth{0308167}\\LPTHE-03-25}
\author{M. Gutperle\\
Department of Physics and Astronomy, UCLA, Los Angeles, CA 90095, USA
\\{\tt E-mail:
gutperle@physics.ucla.edu}}
\author{B. Pioline\\
LPTHE, Universit\'es Paris VI et VII, 4 place  Jussieu, \\
75252 Paris cedex 05, FRANCE\\{\tt E-mail:
pioline@lpthe.jussieu.fr}}
\abstract{ Four-dimensional colliding plane wave (CPW) solutions
have played an important role in understanding the classical 
non-linearities of Einstein's equations. In this note, we investigate
CPW solutions in $2n+2$--dimensional Einstein gravity 
with a $n+1$-form flux. By using an isomorphism  with the
four-dimensional problem, we construct exact solutions analogous
to the Szekeres vacuum solution in four dimensions. 
The higher-dimensional versions of the Khan-Penrose and 
Bell-Szekeres CPW solutions are studied perturbatively 
in the vicinity of the light-cone.
We find that under small perturbations, a curvature singularity 
is generically produced, leading to both space-like and time-like 
singularities. For $n=4$, our results pertain to the collision of two 
ten-dimensional type IIB Blau - Figueroa o'Farrill - Hull - 
Papadopoulos plane waves.
}

\newcommand{\pa}{\partial}

\newcommand{\p}{\partial}

\newcommand{\nn}{\nonumber}
\newcommand{\eps}{\epsilon}
\newcommand{\Real}{\mathbb{R}}

\def\bea{\begin{eqnarray}}
\def\eea{\end{eqnarray}}
\def\be{\begin{equation}}
\def\ee{\end{equation}}
\def\ba{\begin{align}}
\def\ea{\end{align}}
\def\bse{\begin{subequations}}
\def\ese{\end{subequations}}
\def\bi{\begin{itemize}}
\def\ei{\end{itemize}}

\begin{document}

\maketitle
\section{Introduction}
Gravitational colliding  plane wave (CPW) solutions have received much
attention 
over the years, as a way to bring insight into the non-linearities of
the collision of more realistic gravitational waves. The subject
originated in the work of Khan and Penrose \cite{khan} and has a
vast literature, see \cite{Griffbook} for a review and an exhaustive
list of references.

CPW may  play an
important role in primordial cosmology as a possible seed for large
scale structure formation or even setting the initial conditions in a 
pre-big-bang scenario \cite{Kunze}.
At a more formal level, colliding gravitational plane waves 
offer simple models in which to study the fate of the inner null singularity of
realistic Kerr black holes \cite{ori}, and a
useful approximation to scattering at  
Planckian energies \cite{Dray:1984ha}. They also receive 
a cosmological interpretation
as Gowdy universes \cite{Gowdy:jh}  (i.e. with two commuting Killing vectors).

While most of the CPW studies have taken place in the framework of 
four-dimensional Einstein gravity possibly with an Abelian gauge field, 
it is worthwhile to ask which of these results would continue to apply
to higher dimensional gravity theories, and in particular 
to those which describe the low energy limit of string theory
\cite{Gurses:1995tq,Kehagias:1995ki,
Bozza:2000vk,Gurses:2002ki,Gurses:2003xn}. In particular, 
several maximally supersymmetric
plane wave solutions have been identified in type IIB \cite{Blau:2001ne}
and 11-dimensional supergravity \cite{Kowalski-Glikman:wv}.
It is an interesting problem to study their collisions, 
and whether spacelike singularities
are generically produced in such processes. This may be especially
tractable due to the high degree of symmetry of these backgrounds. 
Notice that in order to set-up the collision, it is customary to
restrict to waves with bounded support along $x^+$, thereby breaking
part of the supersymmetries.

Finally, while exact CPW solutions of Einstein equations or their
supergravity generalizations only pertain to very special initial
conditions, it is important to study the stability of plane waves
under small perturbations: indeed, plane waves usually come in
an infinite dimensional moduli space, corresponding to the 
$x^+$-dependent profiles of the various fields restricted by
a single Einstein equation. A particle or string traveling
along the $x^+$ direction will generically be transmitted through the wave, 
but also be partly reflected back and alter the profile of the outgoing wave. 
This process is usually unaccessible in light-cone quantization
approaches, as it involves the emission of $p^+=0$ states. It is
nevertheless of crucial importance when the background wave presents a 
null singularity, as it may turn it into a timelike or spacelike singularity.
Studying this issue at the level of classical gravity may 
in particular shed light on the singularities observed in the 
closed-string parabolic orbifold \cite{lms,Lawrence:2002aj}.

We start in section 2 by reviewing generalities on colliding
plane waves in four dimensions, as well as several exact solutions
which we will aim at generalizing. In section 3, we propose
an ansatz (eq. \eqref{lineelem} below) for $2n+2$-dimensional CPW solutions, 
which leads to same Einstein equations as in four dimensions,  albeit
different boundary conditions. Using this isomorphism, we construct 
some explicit solutions in $2n+2$ dimensions by upgrading four-dimensional 
solutions. In section 4, we develop a perturbative scheme which allows
us to determine the solution in the vicinity of the  light-cone for
arbitrary boundary conditions. We use it to study the stability of 
gravitational and electromagnetic plane waves to small
counter-propagating perturbations. Finally, we apply our method 
to construct higher-dimensional analogues of the Khan-Penrose
and Bell-Szekeres metrics \footnote{Another analogue of the Bell-Szekeres
metric was constructed recently, but in the context of
higher dimensional Einstein-Maxwell gravity~\cite{Gurses:2003xn}.}.
We close in section 5 with a discussion of our results. The results
of the perturbative computations can be found in Appendix A and B.

\section{Colliding plane waves in four dimensions}

\subsection{Generalities}
Colliding gravitational plane waves in four dimensions have been an
field of intensive study. The metric ansatz compatible with the existence of
two commuting spacelike Killing vectors $\partial_x,\partial_y$ is
given by the Rosen-Szekeres line element,
\be
ds^2 = 2 e^{-M} du dv + e^{-U}\big( e^V \cosh W dx^2- 2 \sinh W dx dy
+ e^{-V}\cosh W dy^2\big), \label{lineelemfd}
\ee
where $M,U,V,W$ are functions of the light-cone coordinates
$u,v$ only. We will restrict our attention to colliding waves with 
aligned polarization, ie $W=0$. In addition, one may allow for
an electromagnetic field,
\be
F = \frac12 \left( dH_1 \wedge dx + dH_2 \wedge dy \right)
\label{felemd}
\ee
where $H_1$ and $H_2$ are functions of $u,v$ only. 

In studying plane wave scattering in flat Minkowski space, one usually assumes
that space is flat ahead of each of the incoming plane fronts, say at 
$u<0$ and $v<0$. Space-time is thus divided
into four sectors: in the past region $P:u<0,v<0$, we have flat Minkowski
space with $M=U=V=W=H_i=0$; the right region  $R: u>0, v<0$ corresponds
to the incoming left-moving plane wave, described by $u$-dependent profiles
$U(u),V(u),$ etc; similarly, the left region $L: u<0,
v>0$ corresponds to the right-moving  plane wave, described by $v$-dependent
profiles $U(v),V(v),$  etc. In the forward region $F:u>0,v>0$,
the two waves start to interact, leading to a metric \eqref{lineelemfd}
depending non-trivially on both of the light-cone coordinates.
The problem is thus to determine the functions $U,V,W,M,H_{i}$
in the forward region, given their values on the characteristics
$u=0$ and $v=0$.

\FIGURE{\hfill
\epsfig{file=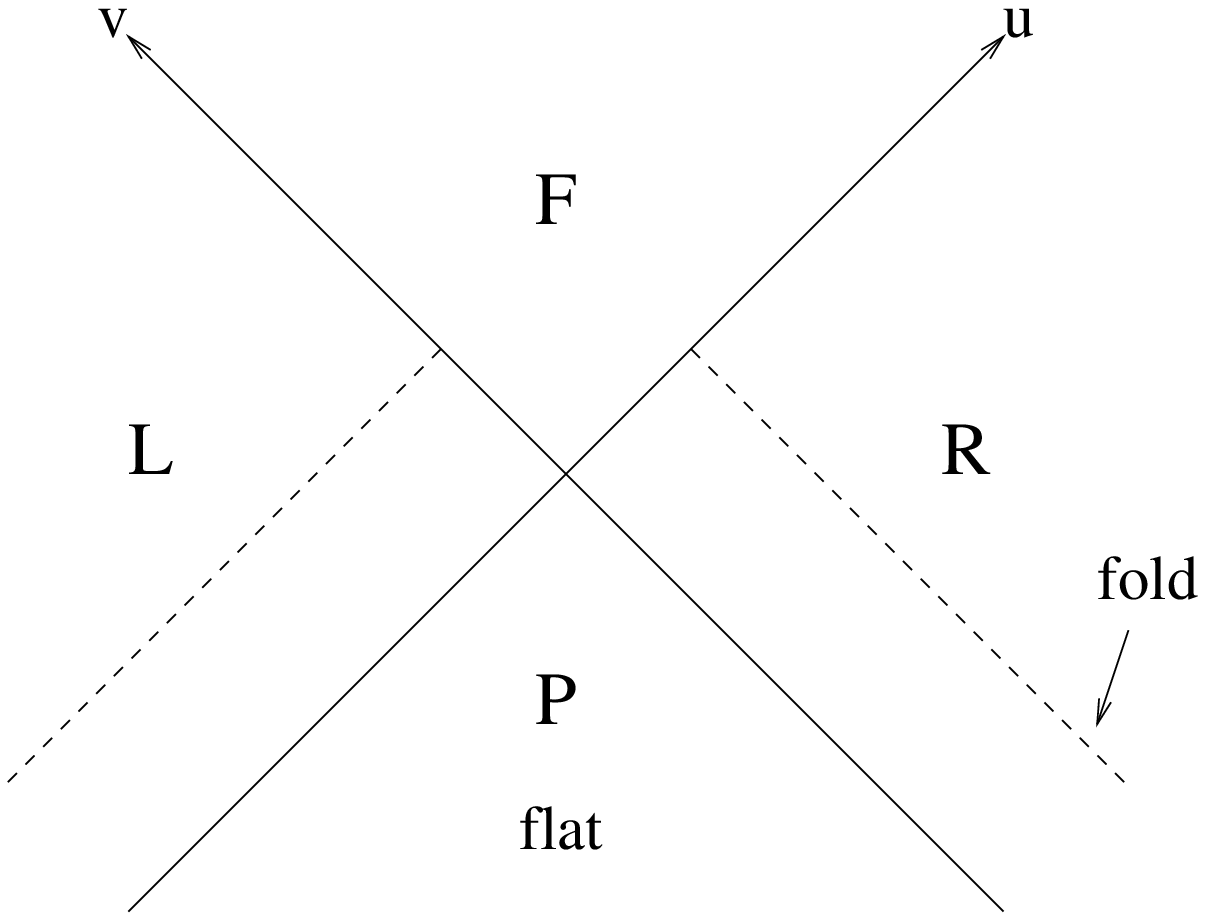,height=8cm}\hfill
\caption{Kinematical set-up  for colliding plane waves. The dotted
line denote a ``fold'' singularity in the incoming regions.}
}

For this, notice that a change of $u$ and $v$ coordinates allows to
set $M=0$ in the right (resp. left) region. The other functions $U,V,W,H_i$
in the incoming region are freely chosen functions of $u$ (resp. $v$), 
subject to the condition that
the corresponding plane wave should satisfy the Einstein equation
$R_{uu}=0$ (resp. $R_{vv}=0$). 
In the interacting region, the Einstein equations require that
$e^{-U}$ is a free two-dimensional field: its value throughout region $F$
is therefore determined immediately in terms of its boundary values
at $u=0$ and $v=0$:
\be
U = -\log \left[ f(u) + g(v) \right]
\label{uprop}
\ee
There is no loss of generality in assuming that $f(0)=g(0)=1/2$.
It is often useful then to change coordinates from $(u,v)$ to $(f,g)$.
Solving for the other functions $V,W,H_i$ is in general a
complicated non-linear problem, except
for purely gravitational  collinear waves  ($W=H_i=0$), where 
$V$ satisfies a linear Euler-Darboux 
equation, and can be determined by Green's function techniques
\cite{Szekeres:uu,Yurtsever:vc}.
When either $W$ or $H_i$ are non-zero, the problem is more difficult,
although integrability provides solution-generating techniques 
which in some cases allow to obtain exact solutions.
Finally, $M$ may be integrated
by quadrature as it satisfies the free 2-dimensional Klein-Gordon equation
with sources (Einstein equations for a higher-dimensional
generalization of the ansatz \eqref{lineelemfd} will be displayed in
Section 3, \eqref{conu}--\eqref{meqatn}). 

While physically realistic waves have a smooth wave front, 
it is often useful to allow for mild singularities at $u=0$ or $v=0$,
in order to describe idealized impulsive or shock wave  profiles.
As shown by O'Brien and Synge \cite{OBrien},
the correct matching conditions to impose at $u=0$ 
are that the transverse  metric
$g_{ij}$ and its derivative $\pa_u  g_{ij}$ be continuous. In terms
of the ansatz  \eqref{lineelemfd},  this implies
that $M$ and $V$ have to be continuous and $f(u)$ and $g(v)$  in
(\ref{uprop}) are at least $C^1$ across the boundaries $u=0$ and
$v=0$. In general then, 
\be
f(u)-{1\over 2} \sim u^{\alpha} \theta(u), \quad g(v)-{1\over
  2} \sim -v^{\beta}\theta(v), \label{fandgdef} 
\ee 
where $\theta(x)$ is the Heaviside step-function and $\alpha, \beta\ge
2$. For $\alpha=2$, there are delta functions in  
components of the curvature tensors which have to 
be interpreted as distributions \cite{Geroch:qn}. Upon writing
the incoming wave in Brinkmann coordinates,
\be
ds^2 =   2 dx^+dx^- + (H_x(x^+) X_i^2 + H_y(x^+)Y_i^2)(dx^+)^2 
+ dX_i^2 +dY_i^2,
\ee
it is easy to see that this corresponds to an impulsive plane wave, 
with $H_{x,y}\propto \delta(x^+)$. The $\alpha=4$ case on the
other hand corresponds to a shock wave, with $H_{x,y}\propto \theta(x^+)$.

\subsection{Exact four-dimensional CPW solutions}
Using the framework just outlined, many exact CPW solutions in
four dimensions have been constructed over the years. We now briefly
review several interesting solutions that we will be interested in
generalizing to higher dimensions, see \cite{Griffbook} for an 
exhaustive review.

A two-parameter family of purely gravitational CPW solutions is given by  
the Szekeres solution \cite{Szekeres:uu}
 \bea
V&=& -2 k_1\;{\rm arctanh}\Big[\sqrt{ {1\over 2}-f\over {1\over 2}+g
}\Big]-
 2 k_2\;{\rm arctanh}\Big[\sqrt{ {1\over 2}-g\over {1\over 2}+f
} \Big],\\
W&=&H=0,\\
M&=&  +{1\over 2 } (1  -(k_1+k_2)^2)\ln
(f+g)  + {k_2^2\over 2} \ln \left({1\over 2}+f \right) 
+{k_1^2\over 2} \ln\left({1\over 2}+g\right)\nonumber \\
&& - 2 k_1k_2\ln\left( \sqrt{{1\over 2}-f}\sqrt{{1\over 2}-g}+
\sqrt{{1\over 2}+f}\sqrt{{1\over 2}+g} \right),\label{mfuncb}  
\eea
where $f=\frac12 - u^\alpha$ and $g=\frac12 - v^\beta$ as in (\ref{fandgdef}) 
and $k_1^2= 2(\alpha-1)/\alpha$, $k_2^2= 2 (\beta-1)/\beta$. A case
of particular interest is the Khan-Penrose solution \cite{khan},
which arises for $\alpha=\beta=2$. In the incoming region $R$,
it corresponds to a profile
\be
\label{kppro}
ds^2\vert_{R} = 2dudv + [1+u\theta(u)]^2 dx^2 + [1-u\theta(u)]^2 dy^2
\ee
which reduces to flat space both before ($u<0$) and after ($u>0$) the 
wave front. The singularity at $u=0$ describes an impulsive gravitational 
wave, with delta function profile. The singularity at $u=-1$ on the other
hand is merely a coordinate singularity, often called ``fold singularity''
in the literature. The profile in the region $L$ is identical
to \eqref{kppro} up to exchanging $u$ and $v$. In the interacting region
however, the geometry is curved, with a space-like curvature singularity 
at $u^2+v^2=1$. The metric furthermore becomes complex at  $u>1$ or
$v>1$. It is thus legitimate to excise the region behind the 
space-like singularity in region $F$, as well as behind the fold singularity
in regions $L,R$. Indeed all but exceptional $u=cste$ causal geodesics
coming from region $P$ fall into the singularity in $F$ \cite{Matzner:pe}.

Another interesting explicit solution is the Bell-Szekeres 
solution \cite{Bell:vb}, which describes colliding
electromagnetic waves in Einstein-Maxwell gravity. The solution in the
interaction region takes the very simple form
given by 
\bea
ds^2&=& 2 du dv + \cos^2(u-v) dx^2+  \cos^2(u+v) dy^2, \label{bsfourd}\\
A_y&=&{1\over \sqrt{2 k}} \sin(u+v) \label{bsfourdb}
\eea
where the replacements $u\to u\theta(u)$ and $v\to v\theta(v)$ are implicit.
In region $R$, the incoming wave may be written in Brinkmann coordinates,
\be
\label{bfhp4}
ds^2 = 2 dX^+ dX^- + dX^2 + dY^2 -\frac14 (X^2+Y^2) (dX^+)^2
\ee
which we recognize the standard electromagnetic plane wave,
One of the most interesting
properties of the Bell-Szekeres solution is that the metric 
\label{bsfourd} in the interaction region $F$ is in fact diffeomorphic 
to a slice of the Bertotti-Robinson $AdS_2\times S^2$ space
\cite{Clarke,Feinstein:2000ja}. 
In contrast of the purely gravitational collision described above, 
the collision of two collinear elecromagnetic waves 
is therefore free of any singularity. Instead, the Killing vectors 
$\pa_x$ and $\pa_y$ become null at $u+v=(2k+1)\pi/2$ and $u-v=
(2l+1)\pi/2$, respectively. There is therefore a Killing horizon
at $u+v=\pi/2$, joining on to the fold singularities at $u=\pi/2$
and $v=\pi/2$ in the incoming regions. In the literature, one usually
excises the region behind the fold singularity and the Killing horizon,
although this appears to be much less justified than in the Khan-Penrose-
Szekeres case.

Note that the incoming plane wave \eqref{bfhp4} is 
the four dimensional analogue of the 
ten dimensional type IIB plane wave found recently by 
Blau - Figueroa o'Farrill - Hull - Papadopoulos (BFHP) \cite{Blau:2001ne}.
One of the goals of this note is to
analyze the scattering of such PP-waves in higher dimensions, a
situation which is of interest in string theory. In order to
avoid undue suspense, let us immediately state that in
contrast to the Bertotti-Robinson space, $AdS_5 \times S^5$ 
cannot be viewed as the collision of two BFHP plane waves.

Finally, we would like to stress that there are two ingredients in
the dynamics of colliding plane waves: firstly the equations
of motions and secondly the matching of the solution in the
interaction region $F$ to incoming plane waves in region $L$ and $R$
and flat space in region $P$. Different matching prescriptions define
different (un)physical problems. In particular, Gowdy-type cosmologies satisfy
the same equations of motion (usually displayed using different
coordinates) but different boundary conditions.

\section{Colliding plane waves in arbitrary dimensions}
Having reviewed the basic features and simplest CPW solutions 
of Einstein-Maxwell gravity in four dimensions, we now generalize
these results to gravity in even dimension $D=2n+2$ with a minimally
coupled  $n+1$ form field strength, whose action is given by
\be
S= \int d^{D} x \sqrt{-g}\Big(R -{k\over 2 (n+1)!} F_{\mu_1 \cdots
  \mu_{n+1}} F^{\mu_1 \cdots  \mu_{n+1}}\Big).
\ee
Our main interest actually lies in the $n=4$ case, which describes
a subsector of ten-dimensional type IIB supergravity, upon restricting to
self-dual configurations of the 5-form field strength. Other cases $n=2,3$ may
also be relevant in the context of type IIB string theory and F-theory
compactified on $K_3$, respectively. This motivates the 
following ans\"atze for the metric,
\be
ds^2 = 2 e^{-M} du dv + e^{-{1\over n}(U-V)} (dx_1^2+\cdots +dx_n^2) +
e^{-{1\over n}(U+V)} (dy_1^2+\cdots +dy_n^2).\label{lineelem} 
\ee
and for the $n+1$-index anti-symmetric tensor field strength,
\bea
F_{n+1} &=& {1\over \sqrt{n}}\Big(\partial_u H_1du\wedge dx_1\wedge
\cdots \wedge dx_n+ 
\partial_v H_1 dv\wedge dx_1\wedge \cdots \wedge dx_n\nonumber\\
&&+ \partial_u H_2 du\wedge dy_1\wedge \cdots \wedge dy_n+
\partial_v H_2 dv\wedge dy_1\wedge \cdots \wedge dy_n\Big).
\label{felem}
\eea
Indeed, these ans\"atze are appropriate for incoming BFHP waves 
in type IIB supergravity \cite{Blau:2002mw} (or their analogues in lower
dimension), which preserve an $SO(n) \times SO(n)$ symmetry.  
Here $M,U,V,H_1$ and $H_2$ are all functions of $u$ and $v$ only.
Factors of $n$ have been inserted for later convenience. One may
also consider non-collinear polarization, or 
switch on other fields present in supergravity such as the
dilaton, axion or other antisymmetric tensor fields, but 
we shall refrain from doing so. Our excuse is that, besides simplicity,
the equations of motion satisfied by the ansatz 
\eqref{lineelem}-\eqref{felem} are in fact identical to those
satisfied by the four-dimensional Rosen-Szekeres
solutions \eqref{lineelemfd}-\eqref{felemd}\footnote{This would
remain true if we allowed non-parallel polarizations preserving
the  $SO(n) \times SO(n)$ symmetry.}.

\subsection{Equations of motion}
Indeed, just as in four dimensions \cite{Szekeres:uu,Griffbook}, 
the Einstein equations on the ansatz \eqref{lineelem}
can be written as two chiral equations,
\bea
\p_u^2 U + \p_u U \p_u M - \frac{1}{2n} \left( (\p_u U)^2 + (\p_u V)^2 \right)
&=&\frac{k}{2 n} \left( e^{U-V} (\p_u H_1)^2+e^{U+V}(\p_u
H_2)^2\right) \label{conu}\\ 
\p_v^2 U + \p_v U \p_v M - \frac{1}{2n} \left( (\p_v U)^2 + (\p_v V)^2 \right)
&=&\frac{k}{2 n} \left( e^{U-V} (\p_v H_1)^2+ e^{U+V}(\p_v
H_2)^2\right) \label{conv} 
\eea 
and three integrability conditions,
\bea
\partial_u\partial_v U- \partial_u U \partial_v U&=&0,\label{ueqan}\\
2\partial_u \partial_v V - \partial_u V\partial_v U  - \partial_v
V\partial_u U + 2 k  \left(e^{U-V}  \partial_u H_1\partial_v H_1-e^{U+V}
\partial_u H_2\partial_v H_2\right)&=&0 \label{ueqcn},\\ 
\partial_u\partial_v M-{1\over 2n} \partial_u V \partial_v V+{2n-1 \over
2n}\partial_u U \partial_v U &=&0. \label{meqatn} 
\eea
Satisfying these equations automatically imply (\ref{conu}) and (\ref{conv}).
In the $L$ and $R$ regions, the latter are simply the equations
$R_{uu}=0$ and $R_{vv}=0$ satisfied by the incoming plane waves.
Note that the constraint (\ref{conu}) (resp. \ref{conv}) 
implies that the mean radius of the transverse
directions $e^{-{U\over 2n}}$ is a concave function of $u$ (resp. $v$)
except when it vanishes. $f(u)$ (resp. $g(v)$) is therefore a concave
decreasing function, indicating that a (possibly coordinate) singularity
in regions $L$ amd $R$ is inevitable. Thus ``fold'' singularities are
generic in higher dimensions as well. Note  also that these equations
automatically imply the vanishing of the Ricci scalar 
\be
R=-\frac1n e^M \left[ (2n+1)\partial_u U \partial_v U
+\partial_u V \partial_v V-2n 
(\partial_u\partial_v M+ 2\partial_u \partial_v V) \right]
\ee
Other curvature invariants $R_2=(R_{\mu\nu})^2$ and 
$R_4=(R_{\mu\nu\rho\sigma})^2$ are however non-trivial, and as we shall
see generically have singularities in the interacting region.
Finally, it is easy to check that that the only conformally flat
plane-wave solution satisfying the ansatz \eqref{lineelem}--\eqref{felem}
is flat $d+2$-dimensional Minkowski space, with the sole exception
of the Bertotti-Robinson metric $AdS_2 \times S^2$ in 4 dimensions. 
In particular, $AdS_5 \times S^5$ does not fit into the ansatz
\eqref{lineelem}--\eqref{felem}in 10 dimensions. 

In addition  to the Einstein equations \eqref{conu}--\eqref{meqatn}, 
the equations of 
motion for the $n$-form electromagnetic gauge field are given by
\bea
2 \partial_u \partial_v H_1 - \partial_u H_1 \partial_v V - \partial_v
H_1 \partial_u V&=&0, \label{feom}\\
2 \partial_u \partial_v H_2 + \partial_u H_2 \partial_v V + \partial_v
H_2 \partial_u V&=&0.
\eea
In dimension $d=6,10$ one can impose a self-duality of the
fieldstrength $F=*F$  which relates the two functions $H_1$ 
and $H_2$ by
\be 
\partial_u H_2 = e^{-V} \partial_u H_1,\quad \partial_v H_2 = -e^{-V}
\partial_v H_1.\label{selfdc} 
\ee
This is notably the case in type IIB supergravity.
In dimensions $d=4,8$ there exist no self-duality condition
in Lorentzian signature, nevertheless (\ref{selfdc}) may still
be imposed. For simplicity, we shall impose (\ref{selfdc})
for all values of $n$, and  denote $H=H_1$. The constraints
\eqref{conu}-\eqref{conv} thus reduce to
\be
\partial^2 U +\partial U\partial M -{1\over 2 n}
\Big((\partial U)^2 +(\partial V)^2\Big)= {k\over n} e^{U-V}(\partial H)^2,
\ee
where $\partial$ stands either for  $\partial_u$ or $\partial_v$.

 As in four dimensions, (\ref{ueqan}) implies that  $e^{-U(u,v)}$ is
a free field in two dimensions, and therefore
\be
U=-\log\big( f(u)+g(v)\big). \label{uprop}
\ee
The functions $f(u)$ and $g(v)$ can be directly determined from
the value of $U$ on the characteristics $u=0$ and $v=0$, assuming
$f(0)=g(0)=1/2$. 
For purely gravitational waves ($H=0$), 
Eq.  \eqref{ueqcn} is the  linear Euler-Darboux equation, and
can be solved by the same Green's function techniques that applied
in the four-dimensional case. When $H\neq 0$, the equations
\eqref{ueqcn} and \eqref{feom} are non-linear, but identical to 
the four-dimensional case: they can thus be dealt with just the
same integrability techniques. The constraints \eqref{conu}-
\eqref{conv} can then be integrated to yield the remaining function $M$.
Changing coordinates from $u,v$ to $f(u),g(v)$, they reduce to
\bea
{\partial_f M}+{f+g \over 2n} (\partial_f V)^2 + {k\over n
}e^{-V} (\partial_f H)^2 -{2n-1\over 2n} {1\over f+g} +
{\partial_u^2 f \over (\partial_uf)^2}&=&0, \\
{\partial_g M}+{f+g \over 2n} (\partial_g V)^2 + {k\over n
}e^{-V} (\partial_g H)^2 -{2n-1\over 2n} {1\over (f+g)} +
{\partial_v^2 g \over (\partial_v g )^2}&=&0.
\eea
As in 4 dimensions, the inhomogeneous terms may be absorbed by defining
\be
M(u,v)= {2n-1\over 2n} \log(f+g) -\log (\partial_u f)-\log (\partial_v g) +
S(f(u),g(v)).\label{newsdef} 
\ee
The constraint equations then simplify to 
\bea
\partial_f S +{1\over n }\Big( k\; e^{-V} (\partial_f H)^2 +{f+g\over  2}
(\partial_f V)^2\Big)&=&0,\\
\partial_g S +{1\over n }\Big( k\; e^{-V} (\partial_g H)^2 +{f+g\over  2}
(\partial_g V)^2\Big)&=&0,
\eea
involving derivatives with respect to  $f,g$ only.

This shows that the equations of motion for the  $2n+2$-dimensional
ansatz \eqref{lineelem}-\eqref{felem} with ``self-dual'' $n+1$ field-strength
are isomorphic to those for the $n=1$ four-dimensional collinear ansatz 
\eqref{lineelemfd}-\eqref{felemd}, upon identifying\footnote{This
isomorphism  remains valid upon dropping the assumptions
of collinearity and self-duality.}
\bea
U_n&=&U_1,\quad  V_n=V_1,\quad  H_n =H_1,\\
M_n &=& {1\over n} M_1+{n-1\over n} \big(\log{(f+g)} - \log(\partial_u f)-
\log(\partial_v g)\big).\label{oneton}
\eea 
This immediately allows us to construct $2n+2$-dimensional
solutions from four dimensional ones, as we now discuss.

\subsection{Higher dimensional CPW from four-dimensions }
As discussed in the previous section the form of the solution has a
very simple dependence on $n$. In particular in the ``self-dual'' case
the equations for $U$ and $V$ are independent of $n$. This implies
that a CPW wave solution in $d=4$ $(n=1)$ automatically gives a
solution in $d=2n+2$, with $n>1$. A word of caution however is that,
due to the non-homogeneous terms $\log(\partial_u f)$,
$\log(\partial_v g)$ in (\ref{oneton}),
the properties of the resulting solution can 
change significantly for $n>1$.  In particular the existence of a
smooth matching of the solution for $n=1$ 
does not in general imply a smooth solution in higher dimensions. In
the following we will analyze the properties of higher dimensional
solution for some particular four dimensional solutions.

\subsubsection*{Purely gravitational higher dimensional CPW}
To illustrate this construction, we start with the four-dimensional
Szekeres  solution \eqref{mfuncb} with 
\be
f(u)= {1\over 2} - u^\alpha\theta(u), \quad 
g(v)= {1\over 2}-
v^\beta\theta(v)  
\ee
Using the upgrading procedure outlined above, one finds a four-parameter
family of purely gravitational CPW solutions in arbitrary dimension $2n+2$
\bea
U&=&-\ln(1- u^{\alpha} -v^{\beta}),\\
V&=& -2 k_1\;{\rm arctanh}\Big[ \sqrt{u^\alpha\over 1-v^\beta}\Big]-
 2 k_2\;{\rm arctanh}\Big[ \sqrt{v^\beta\over 1-u^\alpha}\Big],\\
M&=&   -\ln \partial_u f +{k_1^2\over 2n}\log (1/2-f)  -\ln \partial_v
g +{k_2^2\over 2n}\log (1/2-g)\nonumber \\ 
&+& {2n-1  -(k_1+k_2)^2\over 2n}\ln
(1-u^\alpha-v^\beta) + {k_2^2\over 2n} \ln(1-u^\alpha)+
 {k_1^2\over 2n} \ln(1-v^\beta)
\nonumber \\&+&
  {k_1k_2\over n}\ln( 1- u^\alpha-v^\beta+2
u^\alpha v^\beta +2 \sqrt{u^\alpha v^\beta
(1-u^\alpha)(1-v^\beta)}\Big).\label{mfuncbf} 
\eea
where we implicitely replace $u^\alpha\to u^\alpha
\theta(u)$ and $v^\beta \to v^\beta \theta(v)$. 
As in four-dimensions however, the two parameters $k_1,k_2$ get related
to $\alpha,\beta$ by demanding that $M$ be continuous  across $u=0$ and $v=0$, 
\be
k_1^2 = 2n {\alpha-1\over \alpha}, \quad k_2^2 =2n\ ,
{\beta-1\over \beta}. 
\ee
The family of physical solutions relevant in dimension $2n+2$ is therefore
different from that relevant in dimension 4. The fact that a 
curvature singularity arises at $u\alpha+u^\beta=1$ remains 
nevertheless true. The solution satisfies the appropriate junction
conditions across the null surface $u=0$ (resp. $v=0$)
if $\alpha>2$ (resp. $\beta>2$).  In region $R$, the
incoming plane wave metric is given by 
\bea
ds^2 &=& \big(1-u^\alpha \theta(u)\big)^{- {2n-\alpha\over 2n \alpha}}
\;2 du dv + 
\big(1-u^\alpha\theta(u)\big)^{1\over n} 
 \left({ 1+u^{\alpha\over 2} \theta(u)\over 1- u^{\alpha\over
     2}\theta(u)}\right)^{\sqrt{2(\alpha-1)\over n \alpha}} dx_i^2 \nonumber \\
 &&+ \big(1-u^\alpha\theta(u)\big)^{1\over n}
 \left({ 1+u^{\alpha\over 
     2}\theta(u)\over 1- u^{\alpha\over 
     2}\theta(u)}\right)^{-\sqrt{2(\alpha-1)\over n \alpha}} 
dy_i^2. \label{incommet}
\eea
In order to transform this metric in Brinkmann coordinates one has to
redefine the $u$ coordinate for $u>0$ 
\be
 \big(1-u^\alpha \theta(u)\big)^{- {2n-\alpha\over 2n \alpha}} du  =d\tilde u.
\ee
Then the metric can  be brought into the form
\be
ds^2 = 2 d\tilde u dv + e_x(\tilde u)^2 dx_i^2 + e_y^2(\tilde u)dy_i^2
\ee
and the standard change of variables into Brinkmann form gives
\be
ds^2 = 2 dx^+dx^- + (H_x(x^+) X_i^2 + H_y(x^+)Y_i^2)(dx^+)^2 + dX_i^2 +dY_i^2,
\ee
where  
\be
H_x = {1 \over e_x} {d^2 e_x\over d \tilde u^2} , \quad H_y = 
{1 \over e_y} {d^2 e_y\over d \tilde u^2}.
\ee

An analogue of the four-dimensional Khan-Penrose solution may be 
obtained by setting
 $\alpha=\beta=2$, i.e. $k_1^2=k_2^2n$: the resulting incoming profile 
\bea
H_x&=& {1\over \sqrt{n}} \delta(x^+) + {n-1\over n^{3\over 2}} {x^+\over
  (1-(x^+)^2)^{ n+1\over n}}\theta(x^+), \\
  H_y&=& -{1\over \sqrt{n}} \delta(x^+) - {n-1\over n^{3\over 2}} {x^+\over
  (1-(x^+)^2)^{ n+1\over n}}\theta(x^+). \label{hresx}
\eea
shows an impulsive (delta function) component,together 
with a non-vanishing tail which depends on $x^+$: this is markedly
different from 4-dimensional case, where space was flat  on
either side of the wave front. We shall return to the true analogue of the
Khan-Penrose solution in Sections 3.3 and 4.2.

Similarly, a higher dimensional analog of the Szekeres solution is
obtained by $k_1=k2=\sqrt{2n  {m-1 \over m}}$  which
gives $\alpha=\beta=m$ with integer $m\ge 3$ and one finds 
\be
H_x=  h_x(x^+)\theta(x^+), \quad H_y= h_y(x^+)\theta(x^+),
\label{brinkhresb} 
\ee
where the behavior of $h_x, h_y$ as $x^+\to 0$ depends on the value of
the integer  $m$. For
$m=3$, $h_x,h_y$ diverge as $x^+\to 0$. For $m=4$,  $h_x, h_y$ has
a discontinuity at $x^+\to 0$, but also exhibit a 
$x^+$ dependent tail: this is in contrast to the 4-dimensional Szekeres
solution,  which corresponded to a true shock wave with constant
Brinkmann parameter on either side of the wave front. 
For $m>4$ the functions  $h_x, h_y$ vanish as $x^+\to 0$.  

To summarize, by upgrading the general Szekeres solution we found
exact purely gravitational $SO(n) \times SO(n)$ symmetric 
collinear CPW in higher dimensions. They describe different incoming
profiles from the 4-dimensional case, yet exhibit the same features
of fold singularities in the $R,L$ region and space-like singularity
in the forward $F$ region.

\subsubsection*{Electromagnetic CPW solutions}
We now upgrade the four dimensional Bell-Szekeres
solution \eqref{bsfourd}--\eqref{bsfourdb} into a 
higher dimensional colliding plane wave solution with flux. 
Solving (\ref{conu}) and (\ref{conv}) determines $M$ to be
\bea
M&=&\frac{n-1}{n}  \log\left({\cos(u-v)\cos(u+v)\over \sin(2
  u)\sin(2v)}\right), \\
H&=& ={1\over \sqrt{2 k}} \sin(u+v)
\eea
hence the metric is of the following form
\be
ds = 2 \left(\frac{\sin(2 u)\sin(2v)}{\cos(u-v)\cos(u+v)}
\right)^{\frac{n-1}{n}} du dv
+ \cos^{\frac{2}{n}}(u+v) (dx_1^2 +\dots) +  \cos^{\frac{2}{n}}(u-v)
(dy_1^2 +\dots).\label{BSmeta}
\ee
Unfortunately, this metric exhibit  a curvature singularity on the
characteristics $u=0$ and $v=0$, as can  be  seen  by computing the
scalar curvature invariant
\be
R_2 = \left( { \cos(u-v) \cos (u+v) \over \sin(
  2u)\sin(2v)}\right)^{ 2(n-1)\over n}  
\ee 
This implies that the solution in the interaction region cannot be
glued smoothly to incoming plane waves in region $L$ and $R$, and is
therefore unphysical. Another way to reach  the same conclusion is
to change   variables to  $\tilde u
=(\sin u)^{2n-1\over n}, \tilde v = (\sin v)^{2n-1\over n}$, so as
to remove the singular part of the $dudv$ term in (\ref{BSmeta}):
then $f= 1/2 - \hat u^{ 2n \over 2n-1}$, $ g=1/2 -
\hat v ^{2n\over 2n-1}$ have  exponents $\alpha,\beta$ smaller than two,
in contradiction with the O'Brien Synge conditions. Finally, note
that the incoming plane waves  in the $L$ and $R$ regions are
markedly different from the BFHP plane wave in higher dimensions, 
which was our motivation to look at the Bell-Szekeres solution 
in the first place. 

\medskip

While this attempt  to  produce a physical  
higher-dimensional CPW solution from the  Bell-Szekeres solution
has failed, there are nevertheless four-dimensional  electromagnetic CPW
solutions which lead to acceptable higher-dimensional ones. For
example, recall that using solution-generating techniques
developped in four dimensions, a purely gravitational CPW solution 
$U_0,V_0,M_0$ may be turned into an electromagnetic one by
the  following transformation:
\bea
&&U=U_0, \quad V=V_0 -2 \ln\big( \cos^2\alpha + \sin^2 \alpha\;
e^{-U_0+V_0}\big) \label{mrota} \\
&& M = M_0   -2 \ln\big( \cos^2\alpha + \sin^2 \alpha\;
e^{-U_0+V_0}\big), \quad H= \sqrt{2\over k} {\sin\alpha\cos\alpha\;
(e^{-U_0+V_0}-1)\over \cos^2 \alpha+\sin^2\alpha \;e^{-U_0+V_0}} \nn
\eea
Starting from the  four-dimensional purely gravitational
Szekeres solution \eqref{mfuncb}, one may thus obtain a
5-parameter family of  higher dimensional CPW solutions with flux.
A special choice of parameters then leads to a solution 
which is regular at the characteristics $u=0$ and $v=0$, although
it displays a  curvature singularity at
$f(u)+g(v)=0$ in the interaction region. 


\medskip

\subsection{Toward higher-dimensional Khan-Penrose and Bell-Szekeres CPW}

The construction in the previous section, based on upgrading known
four-dimensional solutions into higher dimensional ones, has failed to 
produce what deserves to be called higher-dimensional analogs of the
Khan-Penrose and Bell-Szekeres solutions, namely solutions describing
the collision of two purely gravitational impulsive plane waves,
and two purely electromagnetic shock waves, respectively. This must 
simply mean that we failed to impose the appropriate boundary 
conditions on the characteristics.

Indeed, a higher-dimensional analog of the Khan-Penrose solution
should describe the collision of two plane waves with Rosen coordinate
metric
\be
\label{kph}
ds^2 = 2dudv + [1+u\theta(u)]^2 (dx_1^2+\cdots +dx_n^2) +
[1-u\theta(u)]^2 (dy_1^2+\cdots +dy_n^2) \ ,
\ee
flat on either side of the wave front at $u=0$.
The appropriate boundary conditions are therefore
\bea
U(u,v) &=& -\log[ (1-u^2)^n + (1-v^2)^n - 1  ]\\
V(u,0) &=& n \log[ (1+u) / (1-u) ] \\
V(0,v) &=& n \log[ (1+v) / (1-v) ] 
\eea
A solution could be obtained using the  Green function technique
developed in \cite{Szekeres:uu,Yurtsever:vc}, however for $n>1$
the resulting integrals appear to be too difficult.

Similarly,  a higher-dimensional analog of the Bell-Szekeres solution
should describe the collision of two BFHP plane waves with Rosen coordinate
metric and $F$ field
\bea
\label{bsh}
ds^2 &=& 2dudv + \cos^2u~ (dx_1^2+\cdots +dx_n^2 + dy_1^2+\cdots +dy_n^2) \\
H &=& 2 \int_0^u \cos^{2n}u du
\eea 
The appropriate boundary conditions are therefore
\bea
U(u,v) &=& -\log[ \cos^{2n }u+ \cos^{2n} v - 1  ]\\
V(u,0) &=& 0\ ,\quad V(0,v) =0 \\
H(u,0)&=& 2 \int_0^u \cos^{2n}u du \ ,\quad
H(0,v)=2 \int_0^v \cos^{2n}v dv 
\eea
Again, it is not clear how to obtain such a solution with
the known solution-generating techniques in four-dimensions.
In the following we will follow a different approach, using
perturbation theory  around the light-cone.

\section{Perturbative plane wave collisions}
In this section, our aim is to analyze higher dimensional plane wave
collisions in a perturbative expansion around the light-cone -- or,
equivalently, when one of the waves is of much smaller amplitude than
the other.  As an application, we shall obtain approximations to
the higher-dimensional Khan-Penrose and Bell-Szekeres CPW solutions
as defined in section 3.3, in the vicinity of the light-cone. We
will also be able to study generic perturbations of the impulsive
and shock plane waves in higher dimensions. 

\subsection{General set-up}
We consider an incoming left-moving plane wave in $2n+2$ dimensions,
given by the Rosen coordinate metric and flux
\bea
ds^2\vert_{R} &=& 2 du dv + e^{-\frac{1}{n}U^{(0)}(u)} \left[
e^{\frac{1}{n}V^{(0)}(u)} 
(dx_1^2+\cdots +dx_n^2) +
e^{-\frac{1}{n}V^{(0)}(u)} 
(dy_1^2+\cdots +dy_n^2) \right] \nn \\
F_{n+1}&=& \frac12 \pa_u H^{(0)}(u) ~du\wedge (dx_{1\dots n} + dy_{1\dots n})
\eea
Here we took advantage of coordinate reparametrization invariance to
set $M=0$, and restricted to self-dual flux configurations. The 
functions $U^{(0)},V^{(0)},H^{(0)}$ are assumed to satisfy the chiral
equation of motion \eqref{conu}. We also assume that these functions
vanish at $u<0$, and satisfy the appropriate regularity conditions
at $u=0$.

On the other hand, 
the incoming right-moving plane wave is assumed to be of the form
\bea
ds^2\vert_{L} &=& 2 du dv + e^{-\frac{1}{n}U^{(1)}(\eps v)} \left[
e^{\frac{1}{n}V^{(1)}(\eps v)} 
(dx_1^2+\cdots +dx_n^2) +
e^{-\frac{1}{n}V^{(1)}(\eps v)} 
(dy_1^2+\cdots +dy_n^2) \right]  \nn \\
F_{n+1}&=& \frac12 \pa_u H^{(1)}(\eps v) 
~du\wedge (dx_{1\dots n} + dy_{1\dots n})
\eea
where we introduced a parameter $\eps$ which can be viewed as
the relative strength of the two plane waves: upon going to 
Brinkmann coordinates, the Brinkmann mass parameter is proportional
to $\eps$. As for the right-moving wave, we assume that 
$U^{(1)}, V^{(1)}, H^{(1)}$ vanish at $v<0$, and have at most
an impulsive singularity at $v=0$. For simplicity, we also restrict
to collinear polarization.

We shall be interested in a perturbative expansion in $\eps$,
therefore in the vicinity of the characteristic axis $v=0$ in the
forward region $F$. To this purpose, we expand the left-moving wave
profile as
\bea
g^{(1)}(v>0) &=& \frac12 + \frac1{2!} g_2 v^2 + \frac1{3!} g_3 v^3 + \dots\\
V^{(1)}(v>0) &=& v_1 v + \frac1{2!} v_2 v^2 + \frac1{3!} v_3 v^3 + \dots\\
H^{(1)}(v>0) &=& h_1 v +  \frac1{2!} h_2 v^2 + \frac1{3!} h_3 v^3 + \dots
\eea
where $g^{(1)}(v)$ is defined as usual 
by $U^{(1)}(v)=-\log[ g^{(1)}(v) + 1/2 ]$. The vanishing of
$g_1, v_0, h_0$ follows from our assumptions on the singularity at $v=0$.
The chiral equation \eqref{conv} allows to eliminate e.g. 
$g^{(1)}$ in favor of $V^{(1)},H^{(1)}$ order by  order in $v$, e.g.
\bea
g_2&=&-\frac{1}{2n}\left( h_1^2 + {v_1}^2 \right)
\ ,\quad
g_3=\frac{1}{2n} \left( 2{h_1}{h_2} - h_1^2{v_1} + 2{v_1}{v_2} \right)\\
g_4&=&  -\frac{1}{4n^3}
\left[ h_1^4\left( 1 - 2n \right)  
+ {v_1}^4 + 
       4{h_1}n^2\left( {h_3} - 
          2{h_2}{v_1} \right) \right. \\
&&\left. + 
       h_1^2\left( \left( n-2 \right) \left( 2n-1 \right) 
           {v_1}^2 - 2n^2{v_2} \right)  + 
       n\left( -3{v_1}^4 + 
          4n\left( {{h_2}}^2 + {{v_2}}^2 + 
             {v_1}v_3 \right)  \right)  \right]  \nn
\eea
Our goal is now to solve for the solution in the interacting region.
Using the field equation \eqref{conu}, $U$ is determined 
throughout the interacting region by its value on the characteristics $uv=0$:
\be
U(u,v) = - \log \left[ f^{(0)} + g^{(1)}(\eps v) \right]
\ee
where we defined as  usual  $U^{(0)}(u)=-\log[ f^{(0)}(u) + 1/2]$.
The other functions $V,H,M$ in the ansatz \eqref{lineelem}-\eqref{felem}
can be determined order by order in $\eps$ by Taylor expanding
\be
\begin{array}{cccccccccc}
H &=& H^{(0)}(u) &+& \eps v H_1(u) &+& \frac1{2!}\eps^2 v^2 H_2(u)
&+& \frac1{3!}\eps^2 v^2 H_3(u) &+ \dots \\
V &=&0 &+ & \eps v V_1(u) &+& \frac1{2!} \eps^2 v^2 V_2(u)
&+& \frac1{3!} \eps^2 v^3 V_3(u) &+ \dots\\
M &=&0 &+& \eps v M_1(u)  &+&\frac1{2!}\eps^2 v^2 M_2(u)&+& 
\frac1{3!}\eps^3 v^3 M_3(u) &+ \dots
\end{array}
\ee
The validity of this expansion may be checked by self-consistency,
or by looking at the known Khan-Penrose and Bell-Szekeres solution for $n=1$.

\subsection{Higher-dimensional Kahn-Penrose solution}
We now specialize to the case where the incoming right-moving plane wave is
a purely gravitational impulsive profile \eqref{kph}, hence 
\be
U(u,v) = -\log  \left[  (1-u^2)^n + g^{(1)}(\eps v) -\frac12  \right]
\ee
At leading order in $\eps$, the equations  \eqref{ueqan}--\eqref{meqatn}
for $H_1(u),V_1(u),M_1(u)$ are first order
homogeneous linear ODE's,
\bea
H_1' - \frac{n}{1-u^2} H_1 &=& 0 \\
V_1' - \frac{n u}{1-u^2} V_1 &=& 0 \\
M_1' -  \frac{1}{1-u^2} V_1 &=& 0 
\eea
Using the boundary conditions $H_1(0)=h_1,V_1(0)=v_1, M_1(0)=0$, we
obtain
\bea
H_1(u) &=& h_1 ~(1+u)^{n/2}~(1-u)^{-n/2}\\
V_1(u) &=& v_1 ~(1-u^2)^{-n/2} \\
M_1(u) &=& v_1 ~ u ~{}_2F_1\left( \frac12, 1+\frac{n}{2}, \frac32, u^2 \right)
\eea
where the hypergeometric function reduces to a simple algebraic function
when $n$ is odd. The norm of the Killing vector $\pa_{x,y}$ thus becomes
\be
\label{kix}
|\pa_{x,y}|^2 = (1-u)^2 \pm \frac1{n} v_1 (1-u)^2 (1-u^2)^{-n/2} 
(\eps v) + {\cal O}(\eps^2)
\ee
which implies that the Killing horizons are shifted to
\be
\label{kixo}
|\pa_{x,y}|^2 =0 : ~~ \eps v \sim \frac{1}{v_1} n~2^{n/2} \Delta^{n/2}
\ee
where $\Delta=1-u$. This is in fact an upper estimate,  since higher
order terms in \eqref{kix}, if singular at $u=1$, may lead to a
critical exponent higher than $n/2$. 
At this order, the curvature invariants $R_2$ and $R_4$ remain zero however.

At second order in $\eps$, the equations for  $H_2(u),V_2(u),M_2(u)$ 
become first order linear ODE's with a source,
\bea
H_2' - \frac{n}{1-u^2} H_1 &=& \frac12 h_1 v_1 n (1-u)^{-n-1}\\
V_2' - \frac{n u}{1-u^2} V_1 &=& \frac12 \left[ v_1^2 - (n+1)h_1^2 \right]
(1-u^2)^{-n-1} \\
M_2' -  \frac{1}{1-u^2} V_2 &=& -\frac{1}{2n} u (1-u^2)^{-n-1}
\left[ (2n-1) h_1^2 + (n-1) v_1^2 \right] 
\eea
These equations are too complicated to give a general answer for any $n$,
however explicit solutions can be found in Appendix B for the values
$n=1\dots 4$ of interest. The main result here is that the curvature
invariant $R_4$ is now non-zero\footnote{
Since they arise only at order $\eps^2$, we systematically 
rescale $R_2$ and $R_4$ by a factor of $\eps^2$. In this purely
gravitational case the Ricci square $R_2$ remains in fact
zero to all orders in $\eps$. }
and in fact diverges in the interacting region
at $u=1$, as $R_4 \sim \Delta^{n+2}$.

We have computed the
solution up to order $\eps^4$ for $n=1\dots 4$ 
in the special case where the left-moving
($v$) profile is purely gravitational, i.e all $h_i=0$.
It is then consistent to set $H(u,v)=0$ throughout the interaction region. 
This allows us to obtain the curvature invariant $R_4$ to second order
in $\eps$ (see Appendix A for explicit results). We find in general that 
higher order corrections diverge faster. Looking at the zeros of $1/R^4$,
we find that  higher order corrections shift the precise location of the pole.
For example, in the $n=4$ case relevant for type IIB, we find
\be
\label{r410}
R_4^{-1}=\frac{16\Delta^6}{3 v_1^2} -  
\frac{11 v_1 \Delta^4}{6 v_1} \eps v 
+ \frac{41 v_1^2 \Delta^2}{192} (\eps v)^2+ {\cal O}(\eps^3)
\ee
which shows that the pole is shifted to\footnote{This relation in fact
holds for any $n$ to order $\eps^4$.} 
\be
R_4=\infty: ~~ \eps v = {\cal O}(\Delta^{n+2}) 
\ee
Determining the precise coefficient 
would require to resum the series in  \eqref{r410}. In fact, more divergent
terms in higher  order contributions to $R^4$ could even increase the
exponent, so that the relation provides only an upper estimate. 
To fourth order, we find that the critical exponent for the location
of the Killing horizon \eqref{kixo} remains equal to $n+2$.
In contrast to the $n=1$ case however, the numerical coefficients appear
to be different: it is therefore conceivable that the space-like singularity
may be hidden behind a Killing horizon, in agreement with cosmic
censorship. 

Finally, a case of particular interest is when the two colliding waves
have identical impulsive profile: this is the higher dimensional analog
of the Khan-Penrose solution. The solution may be obtained to order $\eps^4$
by setting
\be
v_1=2n\ ,\quad v_2=0\ ,\quad v_3=4n\ ,\quad v_4=0\ ,\quad
\ee
in the formulae in Appendix A.

\subsection{Higher-dimensional Bell-Szekeres solution}
We now consider a purely electromagnetic shock wave as our right-moving
background profile given by \eqref{bsh}. The function $U$ is thus determined
throughout the interaction region by
\be
U(u,v) = -\log  \left[  \cos^{2n}u + g^{(1)}(\eps v) -\frac12  \right]
\ee
For the most part, we shall consider purely gravitational left-moving
perturbations, i.e. $v_i=0$. We shall however relax this assumption in 
the $n=1$ case when discussing the stability of its horizon.

To leading order in $\eps$, the equations \eqref{ueqan}
--\eqref{meqatn} then reduces to the linear system
\be
V_1 = \frac{H_1'}{n \cos^n u}\ ,\qquad H_1^{''}+n^2 H_1 = 0
\ee
Using the boundary conditions $V_1(0)=0, H_1(0)=h_1$, one therefore obtains
\be
H_1(u) = h_1 \cos nu\ ,\quad V_1(u) = - h_1 \frac{\sin nu}{\cos^n u}
\ee
One may check that to this order, the curvature invariants 
$R_2=(R_{\mu\nu})^2$ and $R_4=(R_{\mu\nu\rho\sigma})^2$ 
remain zero. The  length of the Killing vectors 
$\pa_{x,y}$ however is shifted to 
\be
\| \p_{x,y} \| ^2 = e^{-(U\mp V)/n} = \cos^2 u \pm 
  \frac{\sin nu}
{n\cos^{n-2} u} \eps h_1 + {\cal O}(\eps^2) \equiv 0
\ee
For $n=1$, the first order correction vanishes at $u=\pi/2$, hence the horizon
remains to this order at $u=\pi/2$. For higher $n$ however,
the correction term blows up at $u=\pi/2$. Were the second order
$\eps^2$ term not more singular than the $\eps$ order term, the horizon 
would be shifted to
\bea
h_1 \eps v \sim \left( u -\frac{\pi}{2} \right)^{n-1} 
\qquad n~ \mbox{even} \\
h_1 \eps v \sim n \left( u -\frac{\pi}{2} \right)^{n} 
\sim  \qquad n~ \mbox{odd}
\eea
Unlike the Khan-Penrose case, it turns out that higher order term
do change this behaviour, and lead to higher exponents. 

At order $\eps^2$, the equations of motion become linear with a source,
\bea
V_2 &=&\frac{H_2'}{n \cos^n u} +
\frac{3 \cos[(2n-1)u]+\cos[(2n+1)u]}{8 \cos^{2n+1} u} h_1^2 \\
 H_2^{''}+n^2 H_2 &=&  -n
\frac{(n-1) \sin[2nu]+2n\sin[2(n-1)u]}{4 \cos^{n+2} u} h_1^2 \\
M_2' &=&   - \frac{n \sin[(2n-1)u] - (3n-2) \sin u}{4n \cos^{2n+1} u} h_1^2
\eea
These equations are readily integrated for given values of $n$, using
the boundary conditions $H_2(0)=h_2, V_2(0)=M_2(0)=0$. The same pattern 
continues to hold to higher order, allowing us to get the solution to
any order desired. Explicit solutions up to fourth order can be found
in the appendix. We now briefly summarize our results. 

In four dimensions, the collision of two collinear
electromagnetic shock wave appears
to be a rather smooth process. Indeed, when the two waves profiles
are identical, the metric in the interaction region is given by 
the Bell-Szekeres solution, which is diffeomorphic to the 
Bertotti-Robinson or $AdS_2 \times S_2$ metric. It does however
present a Killing horizon for the two Killing vectors $\pa_{x,y}$,
at $u+v=\pi/2$. For more general profiles, we find that the
curvature invariants remain finite,
\be
R_2 = R_4 = 4 h_1^2 ~+~ 8 h_1 h_2 \eps  v 
~+~ (h_1^4 +  4 h_2^2 + 4 h_1 h_3)  (\eps  v)^2  + {\cal O}(\eps^3)
\ee
To order $\eps^4$, the norm of the Killing vectors is given by
\be
\|\p_x\|^2 = u^2 \pm h_1 u \eps v + \frac{h_1^2}{4} (\eps v)^2 +
\frac13 h_1 h_2 (\eps v)^3 \pm \frac{11 h_1^2 h_2}{192 u} (\eps v)^4
+ {\cal O}(\eps^5)
\ee
To leading order, the horizon therefore lies at 
\be
u-\pi/2 = \pm \frac 12 h_1 \eps v 
\ee
As is well known however, the collision of an electromagnetic plane
wave with a gravitational wave does lead to a curvature singularity.
Indeed, if we allow for a non-zero $V^{(1)}=v_1 v +  \frac{12} v_2 v^2 
+ \dots$ perturbation, the curvature invariants become
\be
R_2=R_4 = 4(h_1+v_1 \tan u)^2 -\frac{4 
(h_1 \cos u+ v_1 \sin u)((2h_2 - h_1v_1) \cos u+ 2 v_2 \sin u)}{\cos^2 u}
\eps v + {\cal O}(\eps^2)
\ee
where we calculated but do not show the $\eps^2$ term.
As $u\to \pi/2$, the curvature blows up. The precise location of the 
singularity can be found by expanding $1/R_2$ and keeping the
dominant term at each order,
\be
R_2^{-1} = \frac{\Delta^2}{4 v_1^2} - \frac{v_2 \Delta^2}{2 v_1^3}
\eps v - \frac{3}{32 \Delta^2}  (\eps v)^2 + {\cal O}(\eps^3)
\ee
where we defined $\Delta=u-\pi/2$.
The curvature singularity therefore appears at $v = {\cal O}(\Delta^2)$.
For $h_1 = h_2 = 0$ this agrees with the explicit solution found by
Griffiths \cite{Griffiths:qt}.

In contrast, in 6 and higher dimensions, we observe that singularities
are generically created, even for purely electromagnetic plane waves
($V^{(1)}=0$). To order $\eps^3$, the curvature invariants
have a pole of high order at $u=\pi/2$. The order $\eps^4$ contribution
however is still more singular at $u=\pi/2$, implying that the singular
locus is shifted away from $u=\pi/2$. Similarly, the norm of the Killing 
vectors diverges at higher order in $\eps$, leading to a deviation
from $u=\pi/2$. Using the same technique as
in \eqref{r410}, we may derive the upper estimates for the locus of the
Ricci square singularity, Riemann square singularity and of the
Killing horizon:
\bea
R_2 = \infty &:& \quad  
v = {\cal  O}\left[ \left( u- \frac{\pi}{2} \right)^\alpha \right] \\
R_4 = \infty &:& \quad  
v = {\cal  O}\left[ \left( u- \frac{\pi}{2} \right)^\beta \right] \\
\| \pa_{x,y} \| = 0 &:& \quad  
v = {\cal  O}\left[ \left( u- \frac{\pi}{2} \right)^\gamma \right] 
\eea
Our results may be summarized in the following table of upper
critical exponents for $n=1$ through $n=4$,
\be
\begin{array}{c@{\hspace{5mm}}c@{\hspace{5mm}}c@{\hspace{5mm}}c}
n & \alpha & \beta & \gamma  \\ \hline
1 & n.a. & n.a.  & 1 \\
2 & 2 & 3  & 2 \\
3 & 4 & 3  & 3 \\ 
4 & 4 & 5  & 4 \\ \hline
\end{array}
\ee

\vskip 2mm
Generically, it therefore appears that, in  contrast  to  the 
four-dimensional case, for $n>1$ a line of curvature singularity
is created tangentially to the light-cone at $v=0$ at the fold singularities.
One half of this line corresponds to a space-like singularity, while the
other half is time-like. In order to determine the exact critical exponent
of this singular line (instead of the upper bounds that we have derived), 
one would have to analyze the degree of divergence of the curvature 
invariants at $u=\pi/2$ to all orders in $\eps$. In particular,
in the absence of such an analysis, an essential singularity at
$u=\pi/2$ cannot be ruled out. 
It would be interesting to check whether these conclusions remain
valid in the case of non-collinear ($W \neq 0$) or non-purely
electromagnetic ($V\neq 0$) waves.

\FIGURE{\hfill
\epsfig{file=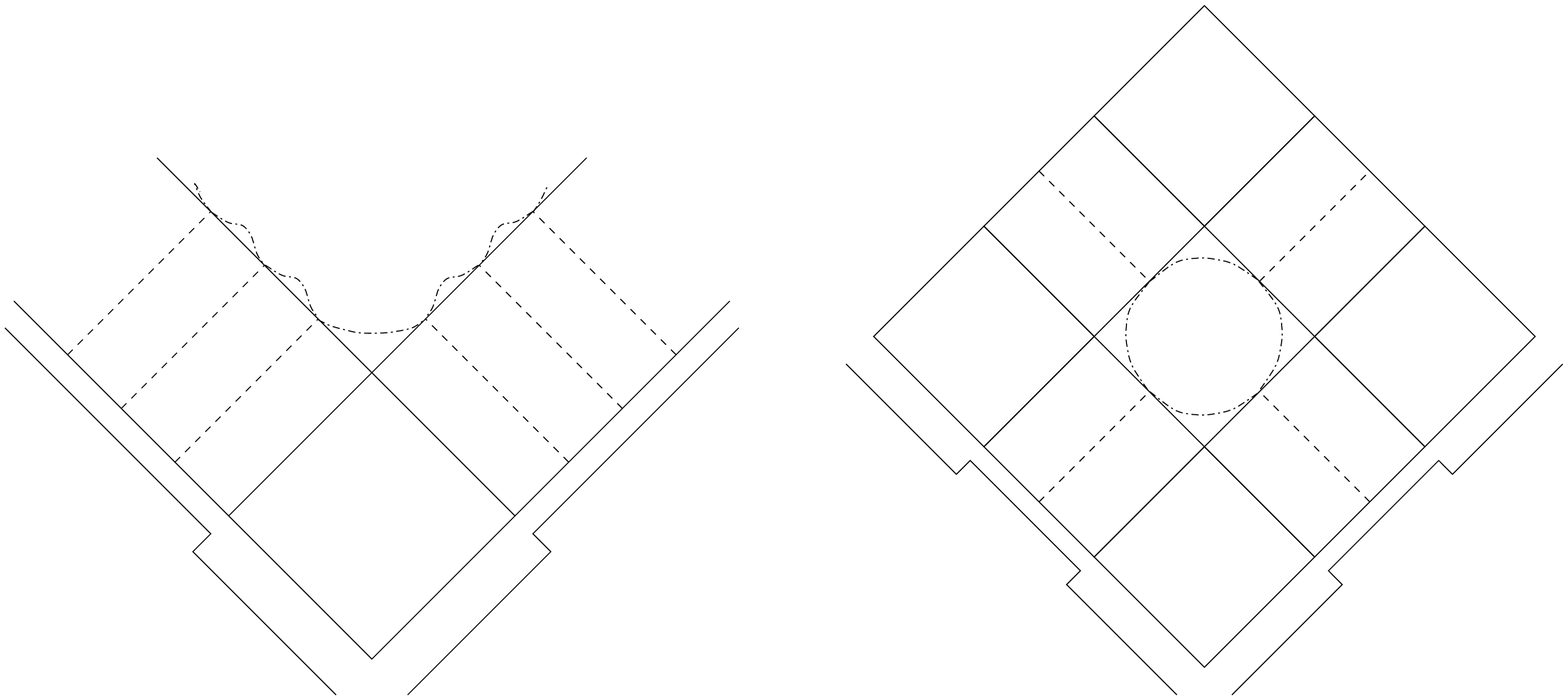,height=7cm}\hfill
\caption{Left: collision of two electromagnetic shock waves for
for $n>1$. Right: collision of two trains of opposite shock waves,
for special choice of pulse duration.}
}

Finally, we may obtain a perturbative solution describing the 
collision of two BFHP shock waves in type IIB supergravity by
specializing our analysis to $n=4$ and 
$H^{(1)}(v)=H^{(0)}(v)$, i.e.
\be
h_1 = 2n\ ,\quad h_2 = 0\ ,\quad h_3 = -2n^2\ , \quad h_4=0\ ,\dots
\ee
The perturbative analysis performed
above then applies in the vicinity of either of the light-cone
axes, with the added simplification that all even coefficients
$h_2, h_4, h_6\dots$ vanish. It is therefore tempting to
conjecture that the spacelike singularity emanating from 
$u=\pi/2, v=0$ joins smoothly onto the one emanating from
$u=0, v=\pi/2$, as depicted in Figure 2. Furthermore, it is plausible that the 
time-like singularity  emanating from  $u=(2p-1)\pi/2$ merges
on to the space-like singularity  emanating from  $u=(2p+1)\pi/2$.
Assuming this is correct, an interesting 
configuration arises when
the incoming waves involves a succession of two 
shock waves of opposite amplitude
separated by a critical time $\pi$: using time reversal invariance,
it is easy to see that the singular locus will form a  closed line
inside the interaction region.

\section{Discussion}
In this note we have studied colliding plane wave solutions of 
$2n+2$-dimensional Einstein gravity with a $n+1$-form field strength. A natural
ansatz \eqref{lineelem}-\eqref{felem} was shown to lead to the same
equations as in ordinary four-dimensional Einstein-Maxwell gravity.
By upgrading known four-dimensional solutions, we were able to
construct exact solutions in $2n+2$ dimensions. In general however,
the boundary conditions suitable in dimension 4 are not appropriate
in higher dimensions. In the second part of the paper, we therefore
developed a perturbative scheme in order to determine CPW solutions with
specified boundary conditions, in the vicinity of the light-cone.

The general conclusion of this study is that, just as in four dimensions,
space-like singularities usually develop in the interaction region, emanating
from the fold singularities in the incoming plane waves. An exception is
the case of four-dimensional purely electromagnetic plane waves, which only
develop a Killing horizon in region $F$. Switching on an arbitrarily small
gravitational perturbation however immediately leads to a space-like
singularity, as was already observed in the context of Kerr black holes
\cite{ori}. In higher dimensions, purely electromagnetic plane wave collisions
already display space-like singularities, as well as Killing horizons. By 
specializing to identical incoming waves, we were able to
obtain the metric of the higher dimensional version of the Khan-Penrose
and Bell-Szekeres CPW solutions in the vicinity of the light-cone.
For $n=4$, the latter describes the collision of two  BFHP plane waves
in type IIB string theory, or rather of their shock wave
generalization.

As in earlier studies of plane wave collisions, it is worth stressing
that the creation of singularities for arbitrary small wave amplitude
is tied to the plane wave symmetry, and would presumably acquire a 
threshold in the case of more realistic pp-waves.

Another general result is that the space-like singularity generally
appears in combination with a time-like singularity, the two of them
meeting at the fold singularity. In earlier studies of the four-dimensional
problem, it has generally be assumed that the region behind the fold
singularity should be excised, thereby getting rid of the time-like
singularity in the forward region. However, in view of recent developments
in the understanding of the asymptotic structure of plane waves
and holography \cite{Marolf:2002ye,Marolf:2003jf,Kiritsis:2002kz,
Berenstein:2002sa}, this prescription seems hard to justify.
Surely, in the absence of the counter-propagating
wave, the fold singularity is simply an artefact of the
Rosen coordinate system. Even in the  presence of the second 
plane wave, there exist null geodesics which can cross
the fold  singularity
and possibly reach the time-like singularity without encountering
any horizon. This does not necessary pose a conflict with cosmic
censorship, as the latter may not apply in non-asymptotically flat
geometries.

In contrast to pp-wave backgrounds, colliding plane wave geometries 
in string theory receive $\alpha'$ and $g_s$ corrections. It would
therefore be very interesting to find exact conformal field theory
descriptions of such geometries. The Bertotti-Robinson $AdS_2
\times S_2$ geometry or the Wess-Zumino-Witten model
$Sl(2)\times SU(2) / \Real \times \Real$ are exact solutions of string
theory \cite{Kehagias:1995ki}, but it is not known how to
impose the junction conditions at $uv=0$ while preserving
conformal invariance. 
On the other hand, string probes propagating in a
pp-wave background are just another example of plane wave collisions,
and it is a very important problem to understand their backreaction
on the plane wave background. 

In this respect,
let us note that there exists a different perturbative
scheme than the one considered in section 4, where one expands 
in the amplitude of the perturbation $g^{(1)}(v)$
rather than its gradient, i.e. define $U(u,v) = - \log \left[ f^{(0)} - 1/2
+ \eps^2 (g^{(1)}(\eps v)-1/2) \right]$. At each order in $\eps$ the profiles
are general functions of $(u,v)$ rather than of $u$ only. Assuming
that the perturbation has compact support along $v$, one may compute
perturbatively its effect on the background right-moving wave. To leading order
in $\eps$, the equations \eqref{ueqan}--\eqref{meqatn} are total derivatives
with respect to $v$; after integration, they become 
ordinary differential equations with respect to $u$ only, with initial
conditions set by the left-moving profile $H^{(1)},V^{(1)}$ at every
value of $v$. This reduction to a one-dimensional dynamical system
is very reminiscent to the dimensional reduction that takes place near
a space-like singularity, however the instability of null singularities
render this observation less useful. 

To conclude, plane wave collisions are an inexhaustible source of
space-like singularities. It would be very useful  to develop a
holographic description of them, as it may provide some insight
into the dynamics of strings near a cosmological singularity.

\acknowledgments 
M.~G. is grateful to the LPTHE, Paris for hospitality during the
early stages of this work. B.~P. is grateful to UCLA Physics
Department and the Aspen Center for Physics for hospitality 
during the final stage of this work. A famous formal computing
software was instrumental in deriving the results presented in section 4.
The work of M.~G. is supported in part by
NSF grant 0245096.
Any opinions, findings and conclusions expressed in this material are
those of the authors and do not necessarily reflect the views of the
National Science Foundation.

\appendix

\section{Perturbative expansions -- the impulsive gravitational case}

\subsection{$n=1$}
At second order,
\bea
H_2(u) &=& h_2 \sqrt\frac{1+u}{1-u}
+h_1 v_1 \frac{1+u - \sqrt{1-u^2}}{2(1-u)}\\
V_2(u) &=&  \frac{u v_1^2}{2(1 - u^2)} + \frac{v_2}{\sqrt{1 - u^2}}
\\
M_2(u) &=&   \frac{u^2 (v_1^2- h_1^2)}{4(1 - u^2)} + 
     \frac{u v_2}{\sqrt{1 - u^2}}
\eea
At third order, for vanishing $h_i$,
\bea
V_3(u) &=&-\frac{2u{\sqrt{1 - u^2}}{v_1}
        {v_2} + u^2\left( {v_1}^3 - 2{v_3} \right)  + 
       2{v_3} }{{2\left( 1 - u^2 \right) }^{\frac{3}{2}}}
\\
M_3(u) &=&   \frac{u\left( \left( 1 + u^2 \right) {v_1}^3 + 
       3u{\sqrt{1 - u^2}}{v_1}{v_2} +
       4\left( 1 - u^2 \right) {v_3} \right) }{4
     {\left( 1 - u^2 \right) }^{\frac{3}{2}}}
\eea
The fourth order correction was computed but is not displayed here.
It allows us to compute the curvature invariant $R_4$ through order $\eps^2$,
\bea
R_4 &=&
\frac{6{v_1}^2}{{\left( 1 - u^2 \right) }^3} + 
  9{v_1}  (\eps v )\frac{3u{\sqrt{1 - u^2}}{v_1}^2 + 
       2{v_2} - 2u^2{v_2}}{{\left( 1 - 
        u^2 \right) }^4} \\
&&+ 3 (\eps v)^2 \frac{
     \left( -1 + 43u^2 \right) v_1^4 + 
       55u{\sqrt{1 - u^2}}v_1^2{v_2} +
       6\left( 1 - u^2 \right) v_2^2 + 
       8\left( 1 - u^2 \right) {v_1}{v_3}}
{2{\left( 1 - u^2 \right) }^4} 
+ {\cal O}(\eps^3) \nn
\eea
Its inverse can be expanded around $u=\pi/2$, keeping the most dominant
term at each order in $\eps$,
\be
R_4^{-1}=-\frac{3v_1^2}{4 \Delta^3} + 
\frac{27 v_1^3}{8 \sqrt{2} \Delta^{7/2}} \eps v 
+ \frac{63 v_1^4}{16 \Delta^4} (\eps v)^2+ {\cal O}(\eps^3)
\ee
The space-like singularity therefore lies at
\be
\eps  v = {\cal O}(\Delta^{1/2})
\ee
Similarly, the length of the Killing vectors reads,  keeping the most dominant
term at each order in $\eps$,
\be
\|\pa_{x}\|^2 = \Delta^2  + \frac{v_1 \Delta^{3/2}}{\sqrt2} \eps v
- \frac{v_2\Delta^{3/2}}{2\sqrt{2}} (\eps v)^2
+\frac{v_1^3 \Delta^{1/2}}{8\sqrt2}  (\eps v)^3
+\frac{v_1^4}{64}  (\eps v)^4
\ee
The Killing horizon thus lies at
\be
\eps  v = {\cal O}(\Delta^{1/2})
\ee

\subsection{$n=2$}
For conciseness, for $n\geq 2$ 
we restrict to the purely gravitational case, $h_i=0$.
At second order in $\eps$,
\bea
V_2(u) &=&\frac{2\left( 4{v_2} + 
        u\left( {v_1}^2 - 4u{v_2} \right)  \right)  
- \left( 1 - u^2 \right) {v_1}^2
      \left( \log\frac{1 - u}{1 + u} \right) }{8{\left( -1 + u^2 \right) }
^2}
\\
M_2(u) &=& \frac{4u
      \left( u{v_1}^2 - 8{v_2} \right)  + 
     4\left( 4{v_2} + u
         \left( {v_1}^2 - 4u{v_2} \right)  \right) 
      \log (\frac{1 - u}{1 + u}) + 
     \left( -1 + u^2 \right) {v_1}^2{\log (\frac{1 - u}{1 + u})}^2}{64
     \left( -1 + u^2 \right) }
\eea
We computed but do not display the third and fourth order contributions.
This allows to compute the curvature invariant $R_4$ through order $\eps^2$,
\bea
R_4 &=& \frac{2(3+u^2) v_1^2}{(1-u^2)^4}
- (\eps v) \frac{v_1 }{4(1-u^2)^5}
  \left[ -2u\left( 31 + 9u^2 \right) 
        v_1^2 \right.\\
&&\left. 
+ 24\left( -3 + 2u^2 + u^4 \right) v_2 + 
       9\left( -3 + 2u^2 + u^4 \right) v_1^2
        \log \left(\frac{1 + u}{1 - u} \right) \right] + {\cal O}(\eps^2)\nn
\eea
Expanding the inverse around $\pi/2$
\be
R_4^{-1}=\frac{2\Delta^4}{v_1^2}+ \frac{5\Delta^3}{2v_1} (\eps v)
+ \frac{87}{64} \Delta^2 (\eps v)^2 + {\cal O}(\eps^3)
\ee
we find that the curvature singularity lies at
\be
\eps v =  {\cal O}(\Delta)
\ee
The length of the Killing vectors reads
\be
\|\pa_{x}\|^2 = \Delta^2  + \frac{v_1 \Delta}{4} \eps v
-\frac{8 v_2 - v_1^2}{64} \Delta (\eps v)^2 - \frac{v_1^3}{384 \Delta}
(\eps v)^3 - \frac{v_1^4}{6144 \Delta^2}
\ee
hence the Killing horizon lies at 
\be
\eps v =  {\cal O}(\Delta)
\ee

\subsection{$n=3$}
At second order,
\bea
V_2(u) &=&\frac{u\left( -3 + 2u^2 \right) {v_1}^2}
    {6{\left( -1 + u^2 \right) }^3} + 
   \frac{{v_2}}{{\left( 1 - u^2 \right) }^{\frac{3}{2}}}
\\
M_2(u) &=& \frac{u \left( u v_1^2 \left( 3 - 6u^2 + 2u^4 \right) 
              + 
       12 v_2 \left( -3 + 2u^2 \right) (1-u^2)^{3/2}  \right) }{36 
        \left( 1 - u^2 \right)^3}
\eea
We computed but do not display the third and fourth order contributions.
This allows to compute the curvature invariant $R_4$ through order $\eps^2$,
\bea
R_4 &=& \frac{2(2 u^2+3)v_1^2}{(1-u^2)^5}
- (\eps v) \frac{v_1}{3(1-u^2)^7}
  \left[ u{\sqrt{1 - u^2}}
          \left( 89 - 36u^4 \right) v_1^2 \right. \\
&&\left.  - 
       18{\left( 1 - u^2 \right) }^2\left( 3 + 2u^2 \right) v_2
       \right]+ {\cal O}(\eps^2) \nn
\eea
Expanding its inverse,
\be
R_4^{-1}=-\frac{16\Delta^5}{5v_1^2}-\frac{10\sqrt{2}\Delta^{7/2}}{3 v_1}
(\eps v) + \frac{309 \Delta^2}{50} (\eps v)^2 + {\cal O}(\eps^3) 
\ee
we find a curvature singularity at
\be
\eps v =  {\cal O}(\Delta^{3/2})
\ee
The length of the Killing vectors reads
\be
\|\pa_{x}\|^2 = \Delta^2 - \frac{v_1 \Delta^{1/2}}{6\sqrt{2}} \eps v
- \frac{v_1^2}{96} (\eps v)^2 -\frac{v_1^3}{5184\sqrt2 \Delta^{5/2}}
(\eps v)^3 +  \frac{v_1^4}{248832\Delta^4} (\eps v)^4
\ee
hence the Killing horizon lies at
\be
\eps v =  {\cal O}(\Delta^{3/2})
\ee

\subsection{$n=4$}
At second order,
\bea
V_2(u) &=&\frac{-2u\left( -5 + 3u^2 \right) {v_1}^2 + 
     32{\left( 1 - u^2 \right) }^2{v_2} + 
     3{\left( 1 - u^2 \right) }^2{v_1}^2\log (\frac{1 + u}{1 - u})}
     {32{\left( 1 - u^2 \right) }^4}
\\
M_2(u) &=&\frac{4u\left( u\left( -23 + 42u^2 - 39u^4 + 12u^6 \right) 
        {v_1}^2 - 32{\left( 1 - u^2 \right) }^2
        \left( -5 + 3u^2 \right) {v_2} \right) }
{1024{\left( 1 - u^2 \right) }^4} \\
&&
 + \frac{
    3
     \left( 4u\left( -5 + 3u^2 \right) v_1^2 - 
       64{\left( 1 - u^2 \right) }^2{v_2} + 
       3{\left( 1 - u^2 \right) }^2 v_1^2
        \left( \log \frac{1 - u}{1 + u} \right)  \right) 
     \log \frac{1 - u}{1+u}  }
{1024{\left( 1 - u^2 \right) }^2} \nn
\eea
We computed but do not display the third and fourth order contributions.
This allows to compute the curvature invariant $R_4$ through order $\eps^2$,
\bea
R_4 &=& \frac{6 v_1^2 (1+u^2)}{(1-u^2)^6}
+\frac{3 v_1 (\eps v)}{16(1-u^2)^8} \left[
-2u\left( -53 - 18u^2 + 27u^4 \right) 
{v_1}^2 \right. \\
&& \left. + { \left( 1 - u^2 \right) }^2\left( 1 + u^2 \right)
\left(
96{v_2} + 27 
       {v_1}^2\log \left(\frac{1 + u}{1 - u} \right) \right)
\right]+ {\cal O}(\eps^2) \nn
\eea
Expanding its inverse,
\be
R_4^{-1}=\frac{16\Delta^6}{3 v_1^2} -  
\frac{11 v_1 \Delta^4}{6 v_1} \eps v 
+ \frac{41 v_1^2 \Delta^2}{192} (\eps v)^2+ {\cal O}(\eps^3)
\ee
we find that the singularity lies at
\be
\eps v = {\cal O}(\Delta^2)
\ee
The length of the Killing vectors reads
\be
\|\pa_{x}\|^2 = \Delta^2 - \frac{v_1}{16} \eps v - \frac{v_1^2}{512 \Delta}
(\eps v)^2  + \frac{v_1^3}{8192 \Delta^3} (\eps v)^3 
+ \frac{9 v_1^4}{2621440\Delta^5}
(\eps v)^4
\ee
hence the Killing horizon lies at
\be
\eps v = {\cal O}(\Delta^2)
\ee

\section{Perturbative expansions -- the electromagnetic case}

\subsection{$n=1$}
At second order, 
\be
H_2(u) = h_2 \cos u - \frac12 h_1^2 \sin u \ ,\quad V_2(u)=-h_2 \tan u\ ,\quad 
M_2(u)=0
\ee
At third order, 
\bea
H_3(u)&=&h_3 \cos u -\frac32 h_1 h_2 \sin u\\
V_3(u)&=& - \frac{\tan u}{8\cos^2 u} 
\left[ 5 h_1^3 +4 h_3 + (h_1^3 + 4 h_3) \cos 2u \right] \\
M_3(u)&=& \frac14 h_1 h_2 \tan^2 u
\eea
At fourth order,
\bea
H_4(u)&=&\frac18 \left[ \left( -5 h_1^2 h_2 + 8{h_4} \right) \cos u 
- \left( 3 h_1^4 + 12 h_2^2 + 
       16h_1{h_3} \right) \sin u+ 5h_1^2h_2\sec u \right]
\\
V_4(u)&=&-\frac{\tan u}{16 \cos^2 u} \left( 45 h_1^2 h_2 + 8{h_4} + 
        \left( 7h_1^2h_2 + 8{h_4} \right) 
         \cos 2u \right) \\
M_4(u)&=& \frac18  \left( h_1^4 + 2h_2^2 + 
      4h_1{h_3} \right) {\tan^2 u}
\eea
The Ricci square reads
\be
R_2 = 4 h_1^2 ~+~ 8 h_1 h_2 \eps  v 
~+~ (h_1^4 +  4 h_2^2 + 4 h_1 h_3)  (\eps  v)^2  + {\cal O}(\eps^3)
\ee
One may check that this is also equal to the Riemann tensor square $R_4=R_2$.

\subsection{$n=2$}
At second order,
\bea
H_2(u)&=&h_2 \cos 2u - h_1^2 \tan u \cos 2u \\
V_2(u)&=& -2 h_2 \tan u\\
M_2(u)&=& h_1^2 \frac{\tan^2 u}{32 \cos^2 u} (5\cos 2u-1)
\eea
At third order, 
\bea
H_3(u)&=& h_3 \cos 2u - h_1 \frac{\left(
      9h_2\left( \cos u + \cos 3u \right)  + 
        2h_1^2\left( 3\sin u - 2\sin 3u \right) 
      \tan (u) \right) }{6 \cos u}\\
V_3(u)&=&\frac16 \left( 4\left( h_1^3 - 3{h_3} \right)  -
        h_1^3{\sec (u)}^2\left( 6 + \sec^2 u \right)  \right)
      \tan u \\
M_3(u)&=& \frac{3h_1h_2\left( 1 + 3\cos (2u) \right) 
    \tan^2 u}{16 \cos^2 u}
\eea
At fourth order, 
\bea
H_4(u)&=& h_4 \cos 2u + \frac{\tan (u)}{12}
\left[ 4\left( 7h_1^4 - 9h_2^2 - 
         12h_1{h_3} \right) \cos (2u)  \right. \\
&& \left.
      + h_1^4\left( -72 + 40{\sec (u)}^2 + {\sec (u)}^4 \right)  + 
      h_1^2h_2\left( -14 + 101\cos (2u) \right) \tan (u)
      \right] \nn \\
V_4(u)&=&\frac{\tan u}{12} \left( 58h_1^2h_2 - 24{h_4} - 
        h_1^2h_2\left( 43 + 36\cos (2u) \right) 
         \sec^4 u \right) \\
M_4(u)&=&-\frac{\sec^2 (u) -1}{1152}
      \left( 529h_1^4 - 1296h_2^2 - 
        1872h_1{h_3} \right. \\
&&\left.
       + \left( -815h_1^4 + 432h_2^2 + 
           432h_1{h_3} \right) {\sec (u)}^2 + 
        h_1^4\left( 179 + 17\cos (2u) \right) {\sec (u)}^6 \right) \nn
\eea
The Ricci square curvature invariant reads
\bea
R_2 &=& 6 \left[ 
(h_1 + h_2 \eps v)^2 + h_1 h_3 (\eps v)^2  \right]  
\frac{\cos^2 2u}{\cos^4 u}  \\
&&\hspace{-5mm} - h_1^4 (\eps v)^2 \frac{\cos 2u}{32 \cos^{6}u}
\left( 1 -90 \cos 2u - 31 \cos 4u \right)\tan^2 u + {\cal O}(\eps^3) \nn
\eea
In order to find the location of the singularity, we expand
$1/R_2$ around $u=\pi/2$, keeping the dominant term at
each order in $\eps$
\be
R_2^{-1}
=\frac{1}{6 h_1^2} \Delta^4 - \frac{h_2}{3 h_1^3}
\Delta^4~\eps v - \frac{5}{96} (v \eps)^2 + {\cal O}(\eps^3)
\ee
At this order, the Ricci curvature singularity therefore lies at
$v = {\cal O}(\Delta^2)$.
As explained in the text, this is only an upper estimate, as higher
order $\eps$-corrections to $R_2$ may be equally or more singular at $u=\pi/2$.
We now turn to the Riemann square curvature invariant,
\bea
R_4&=&
\frac{h_1^2 
\left( 7 + 3\cos (4u) \right) {\sec (u)}^4}{2} \\
&&+ 
  h_1h_2\left( 7 + 3\cos (4u) \right) {\sec (u)}^4
   \eps v \nn\\
&& + \frac{1}{48}
(v \eps)^2 \left[ 12
        \left( -31h_1^4 + 48h_2^2 + 
          48h_1{h_3} \right)  + 
       12\left( 55h_1^4 - 48h_2^2 - 
          48h_1{h_3} \right) {\sec (u)}^2 \right.\nn\\
&& \left.  - 
       5\left( 139h_1^4 - 48h_2^2 - 
          48h_1{h_3} \right) {\sec (u)}^4 + 
       604h_1^4{\sec^6 u} - 263h_1^4{\sec^8u} + 
       90h_1^4{\sec^{10} u} \right] + {\cal O}(\eps^3)\nn
\eea
At each order in $\eps$, the most dominant terms are 
\be
R_4^{-1}=\frac{1}{4 h_1^2} \Delta^4 - \frac{2h_2}{5 h_1^3} \Delta^4
~\eps v - \frac{3}{40 \Delta^2} (\eps v)^2 + {\cal O}(\eps^3)
\ee
The Riemann square singularity is therefore at $v = {\cal O}(\Delta^3)$. 
The length of the Killing vector reads
\be
\| \pa_{x,y} \|^2
= \Delta^2 \pm h_1 \Delta \eps v -\frac{h_1^2}{16 \Delta^2} (\eps v)^2
 \mp \frac{h_1^3}{144 \Delta^3} (\eps v)^3  -\frac{h_1^4}{512 \Delta^6} (\eps v)^4 + {\cal O}(\eps^5)
\ee
The Killing horizons are therefore at $v = {\cal O}(\Delta^2)$.

\subsection{$n=3$}
At second order,
\bea
H_2(u)&=&h_2 \cos 3u -\frac12 h_1^2 \sec^2 u (\sin u -\sin 3u+\sin 5u)\\
V_2(u)&=&-h_2 \frac{\sin 3u}{\cos^3 u} \\
M_2(u)&=&h_1^2 \frac{\tan^2 u}{288 \cos^4 u} 
(-3 + 56 \cos 2u + 43 \cos 4u)
\eea
At third order, 
\bea
H_3(u)&=& h_3 \cos 3u -\frac{h_1 \sin 3u}{240 \cos^5 u}
      \left[ 360h_2{\cos^3u}
\left( 2\cos 2u -1 \right)  \right.\nn\\
&&\left. + 
        h_1^2\left( -80\sin u + 
95\sin 3u - 89\sin 5u
           \right) \right] \nn \\
V_3(u)&=& \frac{1}{60}\tan u \left[464h_1^3 - 240{h_3} + 
      {\sec^2u}\left( -836h_1^3 + 60{h_3} \right) \right.\nn\\
&&\left. + 
         h_1^3{\sec^4u}
          \left( 450 - 103{\sec^2u} + 10{\sec^4u} \right) 
\right]  \nn \\
M_3(u)&=& \frac{h_1h_2
    \left( 33 + 104\cos 2u + 67\cos 4u \right) {\tan^2 u}}
    {144 \cos^4 u}
\eea
At fourth order,
\bea
H_4(u)&=& h_4 \cos 3u + 216 h_1^2 h_2 \cos u + \frac{1}{120}
\left[ \left( -4312h_1^2{h_2} + 120{h_4} \right) \cos 3u 
        - 120{h_4}\cos 4u \right. \nn \\
&&
	- 48\left( 536h_1^4 - 45h_2^2 - 60{h_1}{h_3} \right) \sin u 
	+ 16\left( 176h_1^4 - 45h_2^2 - 60{h_1}{h_3} \right) \sin 3u \nn\\
&&	- 138h_1^4\sec^7u \tan u + 
     {\sec^3u}\left( 8359h_1^2{h_2} -  10396h_1^4\tan u \right)  \nn\\
&&+ 
     {\sec^5u}\left( -807h_1^2{h_2} +  1927h_1^4\tan u \right) \nn \\
&& 
    \left.
- 12\sec u\left( 2430h_1^2{h_2} +  \left( -2156 h_1^4 + 45 h_2^2 + 
           60{h_1}{h_3} \right) \tan u \right) \right] \\
V_4(u)&=&    \frac4{15} \left( 179 h_1^2 h_2  - 15  h_4  \right)  
+ \frac{\tan u}{120 \cos^2 u}
\left[ -10072h_1^2h_2 + 
         120{h_4} \right. \\
&& \left. \hspace*{3cm} +  h_1^2h_2 \left( 5400{\sec^2u}
     - 1181{\sec^4 u} + 95{\sec (u)}^6 \right) \right]\nn \\
M_4(u)&=&\frac{ 1-\sec^2 (u)  }{16200} \left[
4\left( 31049h_1^4 - 15075h_2^2 -  20475h_1{h_3} \right) \right. \\
&&\left.  + 
{\sec^2u} \left( -255964h_1^4 + 36900h_2^2 + 47700h_1{h_3} \right) \right.\nn\\
&&\left.  + 
           {\sec (u)}^4 \left( 178196h_1^4 + 450h_2^2 - 
              900h_1{h_3} \right)\right. \nn \\
&& \left. + 
              9h_1^4{\sec (u)}^6
               \left( -6301 + 1184{\sec (u)}^2 + 50{\sec (u)}^4 \right)
\right] \nn
\eea
The Ricci square reads
\bea
R_2 &=& 8 \left[ 
(h_1 + h_2 \eps v)^2   \right]  \frac{(1-2\cos 2u)^2}{\cos^4 u} 
-\frac{2}{45} h_1 (\eps v)^2 \left( 3{\sec^2u} -4 \right)  
\left[ -2252h_1^3 + 720{h_3} \right. \nn\\
&& \left.+ 
       {\sec^2u} \left( 3849h_1^3 - 540{h_3} \right) + 
          2h_1^3{\sec^4 u}
 \left( -585 - 362{\sec^2 u} + 141{\sec^4u} \right) 
\right] \nn \\
&&+ {\cal O}(\eps^3)
\eea
Keeping the most singular terms at $\pi/2$, we find
\be
R_2^{-1} = \frac{1}{72 h_1^2}
\Delta^4 
+ \frac{47}{6480 \Delta^2} (\eps v)^2 
+ {\cal O}(\eps^3)
\ee
The Ricci singularity is therefore at
$v = {\cal O}(\Delta^3)$.
The Riemann square reads
\bea
R_4&=&\frac{h_1^2}{3 \cos^8 u} 
\left( 8 + 2\cos 2u + 7\cos 4u + 2\cos 6u + 
       \cos 8u \right)\\
&& + \frac{2h_1h_2}{3 \cos^8 u} 
     \left( -\cos 2u + 8 + 10\cos (4u) + 2 \cos 6u+ 
       \cos 8u \right) \eps v \nn \\
&&+ \frac{1}{135} (\eps v)^2 
  \left[ 32\left( -563h_1^4 + 180h_2^2 + 
          180h_1{h_3} \right)  + 
       48\left( 923h_1^4 - 180h_2^2 - 
          180h_1{h_3} \right) {\sec^2 u} \right. \nn \\
&&\left.- 
       6\left( 11261h_1^4 - 1080h_2^2 - 
          1260h_1{h_3} \right) {\sec^4 u} + 
       \left( 92029h_1^4 - 3510h_2^2 - 
          4860h_1{h_3} \right) {\sec^6 u} \right. \nn \\
&&\left.- 
       18\left( 4267h_1^4 - 45h_2^2 - 
          60h_1{h_3} \right) {\sec^8 u} + 
       30690h_1^4{\sec^{10}u} - 
       5040h_1^4{\sec^{12}u} \right. \nn \\
&&\left.+ 
       420h_1^4{\sec^{14}u} \right] + {\cal O}(\eps^3) \nn
\eea
Expanding,
\be
R_4^{-1}= \frac{1}{4 h_1^2} \Delta^8 - \frac 3{h_2}{4 h_1^3}
\Delta^8 ~ \eps v - \frac{7}{36} \Delta^2 (\eps v)^2 + {\cal O}(\eps^3)
\ee
It is singular at $v\sim {\cal O}(\Delta^3)$.
The length of the Killing vector reads
\be
\| \pa_{x,y} \|^2
= \Delta^2 \pm \frac{h_1}{3\Delta} \eps v 
+\frac{h_1^2}{36 \Delta^4} (\eps v)^2
 \pm \frac{h_1^3}{162 \Delta^7} (\eps v)^3  
+\frac{5h_1^4}{3888 \Delta^4} (\eps v)^4 + {\cal O}(\eps^5)
\ee
The Killing horizons are therefore at $v\sim {\cal O}(\Delta^3)$.

\subsection{$n=4$}
At second order,
\bea
H_2(u)&=&h_2 \cos 4u +  2\frac{(\sin u-\sin 3u)\cos 4u}{\cos^3 u} h_1^2 \\
V_2(u)&=& 2\frac{(\sin u-\sin 3u)}{\cos^3 u} h_2 \\
M_2(u)&=& \frac{\tan^2 u (22+439\cos  2u+442 \cos 4u+249 \cos 6u)}
{2048 \cos^6 u} h_1^2
\eea
At third order, 
\bea
H_3(u)&=&\frac{\cos 4u}{32\cos^6 u}
\left( 48h_1^3 + 10{h_3} - 
      3\left( 24h_1^3 - 5{h_3} \right) \cos 2u  + 
      6\left( 8h_1^3 + {h_3} \right) \cos 4u \right.\\
&&\left. - 
      24h_1^3\cos 6u  + {h_3}\cos 6u  - 
      12h_1h_2\sin 2u  - 
      24h_1h_2\sin 4u  - 
      12h_1h_2\sin 6u  \right)  \nn \\
V_3(u)&=&4\left( -2 + \sec^2 u \right) 
  \left( -8h_1^3 + h_3 + 
    h_1^3 \left( 3 + 3\cos 2u + 2\cos 4u \right) \sec^6 u
    \right) \tan u\\
M_3(u)&=& \frac{h_1 h_2}{1024 \cos^6 u} 
    \left( 278 + 823\cos 2u  + 698\cos 4u  + 377\cos 6u  \right) \tan^2 u
\eea
At fourth order,
\bea
H_4(u)&=&h_4 \cos 4u + \frac{\cos 4u}{\cos^3 u} \left(
     3h_2^2\left( \sin u  - \sin 3u  \right)  \right) \\ 
&&+   3h_1^4 \frac{\cos 4u}{\cos^9 u}
   {\left( \sin u  - \sin 3u  \right) }^3 
 + 2h_1{h_3}{\sec^3 u }
   \left( \sin u  - \sin 3u  + \sin 5u  - \sin 7u \right) \nn \\ 
&&+ \frac{h_1^2 h_2 \tan^2 u}{56 \cos^4 u}
     \left( 533 + 72\cos 2u  + 1038\cos 4u  + 8\cos 6u  + 
       505\cos 8u \right) \nn \\
V_4(u)&=&  4{h_4}\tan u \left( \tan ^2 u - 1\right) \\
&&+ \frac{h_1^2  h_2 \tan u }{224 \cos^{10} u}
     \left( 8 - 166\cos 2u  + 64\cos 4u  + 197\cos 6u  + 8\cos 8u + 
       169\cos 10u \right)  \nn \\
M_4(u)&=& h_1{h_3}
\left(\sec^2 u -1 \right) 
\frac{505 - 519\sec^2 u + 121\sec^4 u - 7\sec^6 u}{32} \\
&&+ h_1^4
   \left( \frac{196643}{2048} - 384\sec^2 u + 624\sec^4u - 
     528\sec^6 u + \frac{503857}{2048} \sec^8 u - 60 \sec^{10} u \right.
\nn \\
&& \left. + 
     6\sec^{12}u - \frac{21\sec^{16} u}{512} \right)  + 
  \frac{h_2^2\left( 278 + 823\cos 2u  + 698\cos 4u + 
       377\cos 6u \right) {\tan^2 u}}{1024 \cos^6 u} \nn
\eea
The Ricci square reads
\bea
R_2&=& 10 \left[ 
(h_1 + h_2 \eps v)^2 + h_1 h_3 (\eps v)^2  \right]  
\frac{\cos^2 4u}{\cos^8 u}  \\
&&\hspace{-5mm} -5 h_1^4 (\eps v)^2 \frac{\cos^2 4u}{4096 \cos^{16}u}
\left( -1141 + 1656\cos 2u -  2244\cos 4u + 1992\cos 6u + 
      761 \cos 8u \right) \nn \\
&& + {\cal O}(\eps^3) \nn
\eea
Expanding
\be
R_2^{-1}=
\frac{1}{10 h_1^2} \Delta^8 - \frac{h_2}{5 h_1^3} \Delta^8 ~\eps v
-\frac{49}{640} (\eps v)^2 + {\cal O}(\eps^3)
\ee
It diverges at $v\sim {\cal O}(\Delta^4)$.
The Riemann square reads
\bea
R_4&=&\frac{ 77 - 96\cos (2u) + 48\cos (4u) + 
       5\cos (8u) }{4 \cos^8 u} h_1^2 \\
&& +  
  \frac{ 101 - 144\cos (2u) + 72\cos (4u) + 5\cos (8u)}{2 \cos^8 u} 
h_1h_2 \eps v \nn\\
&& + \frac{1}{128} (\eps v)^2 
     \left[ 320\left( -761h_1^4 + 64h_2^2 + 
          64h_1{h_3} \right)  + 
       640\left( 1273h_1^4 - 64h_2^2 - 
          64h_1{h_3} \right) \sec^2 u \right.\nn\\
&&\left. - 
       16\left[ 104105h_1^4 - 2752h_2^2 - 
          3136h_1{h_3} \right) \sec^4 u \right.\nn\\
&&\left. + 
       16\left( 177121h_1^4 - 2048h_2^2 - 
          2624h_1{h_3} \right) \sec^6 u\right.\nn\\
&&\left. - 
       \left( 3232777h_1^4 - 10304h_2^2 - 
          13376h_1{h_3} \right) \sec^8 u + 
       2103680h_1^4{\sec^{10}u} - 
       714416h_1^4{\sec^{12}u} \right. \nn\\
&&\left. + 
       108848h_1^4{\sec^{14}u} - 
       5083h_1^4{\sec^{16}u} + 
       252h_1^4{\sec^{18}u} \right] \nn
\eea
Expanding
\be
R_4^{-1}=\frac{2}{113 h_1^2} \Delta^8 -\frac{644 h_2}{113^2 h_1^3}\eps v
 -\frac{63}{8\cdot 113^2 \Delta^2} (\eps v)^2 + {\cal O}(\eps^3)
\ee
we see that the $R_4$ curvature invariant is singular at
$v\sim {\cal O}(\Delta^5)$.
The length of the Killing vector reads
\be
\| \pa_{x,y} \|^2
= \Delta^2 \pm \frac{h_1}{\Delta} \eps v 
-\frac{h_1^2}{64 \Delta^6} (\eps v)^2
 \mp \frac{h_1^3}{64 \Delta^9} (\eps v)^3  
-\frac{3h_1^4}{8192 \Delta^4} (\eps v)^4 + {\cal O}(\eps^5)
\ee
The Killing horizons are therefore at $v\sim {\cal O}(\Delta^4)$.


\begin{thebibliography}{00}

\bibitem{khan}
K. A. Kahn and R. Penrose, 
``Scattering of Two Impulsive Gravitational Plane Waves''
Nature {\bf 229} (1971) 185

\bibitem{Griffbook}
J.~B.~Griffiths, "Colliding Plane Waves in General Relativity",
Clarendon Press, 1991




\bibitem{Kunze}
A.~Feinstein, K.~E.~Kunze and M.~A.~Vazquez-Mozo,
``Initial conditions and the structure of the singularity in
pre-big-bang  cosmology,'' 
Class.\ Quant.\ Grav.\  {\bf 17} (2000) 3599
[arXiv:hep-th/0002070].


\bibitem{ori}
A.~Ori and E.~E.~Flanagan,
``How generic are null spacetime singularities?,''
Phys.\ Rev.\ D {\bf 53}, 1754 (1996)
[arXiv:gr-qc/9508066];

L.~M.~Burko and A.~Ori,
``Analytic study of the null singularity inside spherical charged black  holes,''
Phys.\ Rev.\ D {\bf 57}, 7084 (1998)
[arXiv:gr-qc/9711032];

A.~Ori,
``Null weak singularities in plane-symmetric spacetimes,''
Phys.\ Rev.\ D {\bf 57} (1998) 4745
[arXiv:gr-qc/9801086];

A.~Ori,
``Evolution of scalar field perturbations inside a Kerr black hole,''
Phys.\ Rev.\ D {\bf 58} (1998) 084016;

A.~Ori,
``Evolution Of Linear Gravitational And Electromagnetic Perturbations Inside A Kerr Black Hole,''
Phys.\ Rev.\ D {\bf 61} (2000) 024001.



\bibitem{Dray:1984ha}
T.~Dray and G.~'t Hooft,
``The Gravitational Shock Wave Of A Massless Particle,''
Nucl.\ Phys.\ B {\bf 253} (1985) 173.


\bibitem{Gowdy:jh}
R.~H.~Gowdy,
``Gravitational Waves In Closed Universes,''
Phys.\ Rev.\ Lett.\  {\bf 27} (1971) 826.




\bibitem{Gurses:1995tq}
M.~Gurses and E.~Sermutlu,
``Colliding gravitational plane waves in dilaton gravity,''
Phys.\ Rev.\ D {\bf 52}, 809 (1995)
[arXiv:hep-th/9503218].

\bibitem{Kehagias:1995ki}
A.~A.~Kehagias and E.~Papantonopoulos,
``Discontinuities and collision of gravitational waves in string theory,''
Mod.\ Phys.\ Lett.\ A {\bf 10} (1995) 2531
[arXiv:hep-th/9501070].



\bibitem{Bozza:2000vk}
V.~Bozza and G.~Veneziano,
``O(d,d)-invariant collapse/inflation from colliding superstring waves,''
JHEP {\bf 0010} (2000) 035
[arXiv:hep-th/0007159].


\bibitem{Gurses:2002ki}
M.~Gurses and A.~Karasu,
``Some higher dimensional vacuum solutions,''
Class.\ Quant.\ Grav.\  {\bf 18}, 509 (2001)
[arXiv:gr-qc/0007068];
M.~Gurses, E.~O.~Kahya and A.~Karasu,
``Higher dimensional colliding gravitational plane wave metrics,''
Phys.\ Rev.\ D {\bf 66} (2002) 024029
[arXiv:gr-qc/0204041].


\bibitem{Gurses:2003xn}
M.~Gurses, Y.~Ipekoglu, A.~Karasu and C.~Senturk,
``Higher dimensional Bell-Szekeres metric,''
arXiv:gr-qc/0305105.




\bibitem{Blau:2001ne}
M.~Blau, J.~Figueroa-O'Farrill, C.~Hull and G.~Papadopoulos,
``A new maximally supersymmetric background of IIB superstring theory,''
JHEP {\bf 0201} (2002) 047
[arXiv:hep-th/0110242].


\bibitem{Kowalski-Glikman:wv}
J.~Kowalski-Glikman,
``Vacuum States In Supersymmetric Kaluza-Klein Theory,''
Phys.\ Lett.\ B {\bf 134} (1984) 194.



\bibitem{lms} 
H.~Liu, G.~Moore and N.~Seiberg,
``Strings in a time-dependent orbifold,''
JHEP {\bf 0206}, 045 (2002)
[arXiv:hep-th/0204168];
``Strings in time-dependent orbifolds,''
JHEP {\bf 0210}, 031 (2002)
[arXiv:hep-th/0206182].

\bibitem{Lawrence:2002aj}
A.~Lawrence,
``On the instability of 3D null singularities,''
JHEP {\bf 0211} (2002) 019
[arXiv:hep-th/0205288].

\bibitem{OBrien}
S.~O'Brien and J.L.~Synge, ``Jump conditions at discontinuities in
General Relativity'', Comm. Dublin Inst. Adv. Stud. A9 (1952)

\bibitem{Geroch:qn}
R.~Geroch and J.~Traschen,
``Strings And Other Distributional Sources In General Relativity,''
Phys.\ Rev.\ D {\bf 36} (1987) 1017.

\bibitem{Szekeres:uu}
P.~Szekeres,
``Colliding Plane Gravitational Waves,''
J.\ Math.\ Phys.\  {\bf 13} (1972) 286.

\bibitem{Matzner:pe}
R.~A.~Matzner and F.~J.~Tipler,
``Metaphysics Of Colliding Selfgravitating Plane Waves,''
Phys.\ Rev.\ D {\bf 29} (1984) 1575.

\bibitem{Bell:vb}
P.~Bell and P.~Szekeres,
``Interacting Electromagnetic Shock Waves In General Relativity,''
Gen.\ Rel.\ Grav.\  {\bf 5} (1974) 275.


\bibitem{Clarke}
C.~J.~S.~Clarke and S.~A.~Hayward,
``The global structure of the Bell-Szekeres solution'',
Class.\ Quantum Grav.\ {\bf  6} (1989) 615.

\bibitem{Feinstein:2000ja}
A.~Feinstein,
``Penrose limits, the colliding plane wave problem and the classical  string backgrounds,''
Class.\ Quant.\ Grav.\  {\bf 19} (2002) 5353
[arXiv:hep-th/0206052].




\bibitem{Griffiths:qt}
J.~b.~Griffiths,
``Colliding Plane Gravitational And Electromagnetic Waves,''
J.\ Phys.\ A {\bf 16} (1983) 1175.



\bibitem{Yurtsever:vc}
U.~Yurtsever,
``Structure Of The Singularities Produced By Colliding Plane Waves,''
Phys.\ Rev.\ D {\bf 38} (1988) 1706.




\bibitem{Berenstein:2002jq}
D.~Berenstein, J.~M.~Maldacena and H.~Nastase,
``Strings in flat space and pp waves from N = 4 super Yang Mills,''
JHEP {\bf 0204} (2002) 013
[arXiv:hep-th/0202021].

\bibitem{Blau:2002mw}
M.~Blau, J.~Figueroa-O'Farrill and G.~Papadopoulos,
``Penrose limits, supergravity and brane dynamics,''
Class.\ Quant.\ Grav.\  {\bf 19} (2002) 4753
[arXiv:hep-th/0202111].

\bibitem{Blau:2002dy}
M.~Blau, J.~Figueroa-O'Farrill, C.~Hull and G.~Papadopoulos,
``Penrose limits and maximal supersymmetry,''
Class.\ Quant.\ Grav.\  {\bf 19} (2002) L87
[arXiv:hep-th/0201081].




\bibitem{Marolf:2002ye}
D.~Marolf and S.~F.~Ross,
``Plane waves: To infinity and beyond!,''
Class.\ Quant.\ Grav.\  {\bf 19} (2002) 6289
[arXiv:hep-th/0208197].

\bibitem{Marolf:2003jf}
D.~Marolf and S.~F.~Ross,
``Plane waves and spacelike infinity,''
arXiv:hep-th/0303044.

\bibitem{Kiritsis:2002kz}
E.~Kiritsis and B.~Pioline,
``Strings in homogeneous gravitational waves and null holography,''
JHEP {\bf 0208} (2002) 048
[arXiv:hep-th/0204004].


\bibitem{Berenstein:2002sa}
D.~Berenstein and H.~Nastase,
``On lightcone string field theory from super Yang-Mills and holography,''
arXiv:hep-th/0205048.


\end{thebibliography}
\end{document}